\documentclass[preprint,superscriptaddress]{revtex4-1}

\usepackage{latexsym}
\usepackage{natbib}
\usepackage{amsmath}
\usepackage{amsfonts}
\usepackage{amssymb}
\usepackage[utf8]{inputenc}
\usepackage[T1]{fontenc}
\usepackage[dvipdfm]{color,graphicx}
\usepackage{verbatim}
\usepackage{t1enc}
\usepackage{mathrsfs}
\usepackage{tocbibind}
\usepackage[percent]{overpic}
\usepackage{float} 
\usepackage{appendix}
\usepackage{booktabs} 
\usepackage{natbib}
\usepackage{array}
\usepackage{xcolor}
\usepackage{subfigure}

\begin{document}
\title{Nonmonotonic percolation threshold in correlated networks and hypergraphs}


\author{L.  D. Valdez}

\affiliation{Departamento de F\'isica, FCEyN, Universidad Nacional de Mar del Plata, Mar del Plata 7600, Argentina.}
\affiliation{Instituto de Investigaciones F\'isicas de Mar del Plata (IFIMAR), CONICET, Mar del Plata 7600, Argentina.}

\author{C. E. La Rocca}

\affiliation{Departamento de F\'isica, FCEyN, Universidad Nacional de Mar del Plata, Mar del Plata 7600, Argentina.}
\affiliation{Instituto de Investigaciones F\'isicas de Mar del Plata (IFIMAR), CONICET, Mar del Plata 7600, Argentina.}
  \date{\today}

\begin{abstract}
We study the effect of assortative and disassortative mixing on the robustness of networks under random node failures. For ordinary (dyadic) networks, by using the generating function technique and stochastic simulations, we show that the relationship between the Pearson assortativity coefficient $r$ and the percolation threshold $p_c$ is not always monotonic. More specifically, in certain regions of the parameter space of our model, moderately disassortative networks can be more fragile than either strongly disassortative or uncorrelated networks. We observe this nonmonotonic behavior for trimodal networks as well as for networks with Poisson and power-law degree distributions. We then extend our analysis to hypergraphs with correlations between node hyperdegree and hyperedge cardinality. For this case, we find that positively correlated hypergraphs tend to be more fragile than negatively correlated ones. Additionally, as in the dyadic case, the relationship between $r$ and $p_c$ is nonmonotonic, and the most fragile configuration does not correspond to the most assortative hypergraph.
\end{abstract}

\maketitle

\section{Introduction}

Over the last 25 years, research on real-world networks has shown that many systems have structures that differ substantially from the homogeneous and uncorrelated network models used in classical random graph theory. For instance, pioneering work by Barab\'asi and Albert revealed that many real networks have a scale-free degree (or connectivity) distribution, in which the vast majority of nodes have a small degree, while a small but significant number of nodes (called hubs) have a very large number of connections~\cite{albert2002statistical,barabasi1999emergence}.  Additionally, multiple studies have shown that real networks are also characterized by non-trivial degree-degree correlations between neighboring nodes~\cite{boccaletti2006complex,artime2024robustness}. For instance, some online social networks tend to exhibit assortative mixing, which means that nodes are more likely to connect to others with a similar number of connections~\cite{al2023examining,artime2024robustness}. In contrast, biological systems and technological infrastructures usually show disassortative mixing, where highly connected nodes are more likely to connect to low-degree nodes~\cite{al2023examining,artime2024robustness}. These structural patterns have been found to strongly influence a wide range of processes, including spreading dynamics~\cite{hasegawa2012robustness,chang2020impact}, synchronization~\cite{li2013reexamination,roy2021assortativity,la2011synchronization}, and traffic~\cite{sun2009effect}. In this work, we specifically investigate how assortative and disassortative mixing affect the robustness of complex networks under random node failures.

To describe how nodes tend to connect with others based on their degree, researchers have used two main tools: the joint excess-degree distribution and the Pearson assortativity coefficient~\cite{newman2011structure,serrano2007correlations}. The joint excess-degree distribution $P_1(\widetilde{k},\widetilde{k'})$ 
gives the probability that, following a randomly selected link, the two endpoints have $\widetilde{k}$ and $\widetilde{k'}$
outgoing links (excluding the link under consideration). This bivariate distribution gives a detailed description of degree-degree correlations between neighboring nodes, and it has been used in some theoretical studies~\cite{newman2002assortative,newman2003mixing,goltsev2008percolation}. However, due to its high dimensionality, direct comparison of networks using $P_1(\widetilde{k},\widetilde{k'})$ is often impractical. On the other hand, the Pearson assortativity coefficient, usually denoted as $r$, is a scalar quantity derived from $P_1(\widetilde{k},\widetilde{k'})$ that ranges between -1 and 1. Networks with assortative mixing tend to have a positive value of $r$, while those with $r<0$ indicate disassortative mixing, and $r\approx 0$ suggests weak or no correlation. Of course, as has been pointed out in numerous studies, the Pearson coefficient is an imperfect measure because it captures only part of the information contained in $P_1(\widetilde{k},\widetilde{k'})$ and may miss important structural details~\cite{dorogovtsev2010zero,hao2011dichotomy,mussmann2015incorporating}. Nevertheless, this coefficient still remains widely used because it is a single number that allows researchers to quickly compare different networks.

More recently, the concept of degree-degree correlations has been extended beyond ordinary networks to hypergraphs and other higher-order networks~\cite{guzman2026unveiling,sampson2025oscillatory,landry2022hypergraph,ha2025connected,mann2022degree,mann2025mixing}. Unlike ordinary (or dyadic) networks, where edges connect pairs of nodes, hypergraphs allow interactions among groups of nodes through hyperedges.  Numerous studies have shown that hypergraphs are useful for describing social interaction networks, chemical structures, 5G wireless networks, among others~\cite{de2020social,konstantinova2001application,zhang2017hypergraph,bretto2013hypergraph}. However, as recently discussed in Ref.~\cite{peixoto2026graphs}, hypergraphs are not the only representation capable of capturing group or higher-order interactions. In a hypergraph, the hyperdegree of a node is defined as the number of hyperedges it participates in, and the cardinality (or size) of a hyperedge is defined as the number of nodes it connects. Several approaches have been explored to introduce and analyze correlations in hypergraphs. For instance, Sampson et al.~\cite{sampson2025oscillatory} proposed an opinion dynamics model on hypergraphs in which nodes interact via both pairwise links and higher-order hyperedges, and investigated how correlations between these two types of interactions affect the emergence of oscillatory and excitable opinion dynamics. In a different direction, Ha et al.~\cite{ha2025connected}  showed that theoretical models that ignore correlations between node hyperdegree and hyperedge cardinality can, in some cases, overestimate the fraction of nodes belonging to the giant component (GC) in real hypergraphs.

As mentioned above, degree-degree correlations have been shown to influence a wide range of static and dynamical processes on networks~\cite{boccaletti2006complex,roy2021assortativity,noh2007percolation,valdez2011effect}. One of the earliest studies addressing their impact on network robustness was conducted by Newman~\cite{newman2002assortative,newman2003mixing}. Specifically, he investigated a node percolation process on ordinary networks with varying levels of degree-degree correlation. In this process, each node is independently removed (i.e., fails) with probability $1-p$, and the critical threshold $p=p_c$ is defined as the point below which the GC is destroyed. Newman found that assortative networks exhibit a lower percolation threshold $p_c$ than disassortative networks, which suggests that networks become more fragile under random failures when $r$ decreases. However, it is important to note that Newman's work only explored a limited number of networks with specific correlation values, such as $r=-0.20$, $r=0$, and $r=0.20$. A more systematic exploration of the role of degree correlations was later introduced by Mizutaka and Tanizawa~\cite{mizutaka2016robustness}. Specifically, they studied a node percolation process on "bimodal networks" that allowed them to explore the full range of degree correlations, from $r=-1$ to $r=1$. Here, the term bimodal networks refers to networks in which nodes can take only two possible degree values, denoted by $k_1$ and $k_2$, and no other degrees are allowed. For example, if $k_1=3$ and $k_2=5$, each node is connected to either three or five neighbors. Their results demonstrated that the percolation threshold in bimodal networks increases monotonically as the assortativity coefficient decreases. 

However, more recently, Wang et al.~\cite{wang2022basic} showed that the monotonic relation between network robustness and the Pearson assortativity coefficient is not always valid. Although their work focused on epidemic spreading, their conclusions also apply to node and link percolation processes. In particular, they found that disassortative networks can, in some cases, be more robust than networks generated by the configuration model~\cite{molloy1995critical} (with $r=0$). Previous work by Hasegawa et al.~\cite{hasegawa2012robustness} has also shown via stochastic simulations that the GC of susceptible nodes can display nonmonotonic behavior, but the transition point for the final outbreak size (which corresponds to the GC in a percolation process) remains monotonic with respect to $r$. To the best of our knowledge, the study by Wang et al. is the only one challenging the notion that disassortative dyadic networks are necessarily more fragile. On the other hand, while node percolation has been investigated in uncorrelated hypergraphs, extensions to correlated hypergraphs remain largely unexplored.

In this manuscript, we explore a random node percolation process in both ordinary (or dyadic) networks and hypergraphs with assortative and disassortative mixing. For dyadic networks, extending the original work of Mizutaka and Tanizawa, we study a node percolation process in trimodal networks, which are defined as networks where nodes can have one of three possible degree values. Using stochastic simulations and the generating function technique, we find that assortative networks are more robust than disassortative networks, in agreement with previous findings in the literature. Additionally, our results show that, within a certain region of the parameter space of our model, disassortative networks can be more robust than uncorrelated networks, in line with the results reported by Wang et al. However, we also find that for disassortative networks, the percolation threshold $p_c$ can exhibit a nonmonotonic behavior, especially as the correlation approaches $r = -1$, which indicates that the robustness of disassortative networks is far from trivial. We also observe the same behavior in networks with Poisson and power-law degree distributions. On the other hand, for hypergraphs with hyperdegree-cardinality correlations, we find that positively correlated hypergraphs are more fragile than those with negative correlations. Moreover, we show that the percolation threshold $p_c$ exhibits a nonmonotonic behavior for positively correlated networks, particularly in the region where $r$ is close to 1.

This manuscript is organized as follows. In Sec.~\ref{sec.Backg}, we briefly review the theoretical framework for node percolation in correlated networks. In Sec.~\ref{sec.PerCorrDy}, we present our results for trimodal networks and for networks with Poisson and power-law degree distributions. In Sec.~\ref{sec.HyperDegCar}, we extend the analysis to hypergraphs with degree-cardinality correlations, and derive the corresponding self-consistent equations and percolation threshold. Finally, in Sec.~\ref{sec.Concl}, we summarize our main findings.

\section{Background}\label{sec.Backg}
Among all percolation processes, random node percolation is one of the simplest and has been widely used to study how robust a network is against random failures~\cite{stauffer2018introduction,newman2001random,li2021percolation}. Consider a network where a macroscopic fraction $P_{\infty}$ of nodes belongs to the GC, that is, they are part of a globally connected cluster. In a node percolation process, each node is randomly removed (along with its links) with probability $1-p$. At the extreme case of $p=1$, the original network remains intact, and we will say that the network is in a fully functional phase. On the other hand, at the opposite extreme $p=0$, all nodes have been removed and clearly $P_{\infty}=0$.

As we decrease $p$ from $p=1$, the network becomes progressively "diluted" and the relative size of the giant component, $P_{\infty}$, decreases. For networks with an infinite number of nodes, it has been shown that there exists a threshold $p=p_c$ below which $P_{\infty}=0$. 

Callaway et al.~\cite{callaway2000network} showed that for uncorrelated random networks (generated with the configuration model~\cite{molloy1995critical}), both $P_{\infty}(p)$ and the threshold $p_c$ can be calculated exactly using percolation and branching theory. In particular, it was shown that the critical threshold is given by $p_c=1/(\langle k^2\rangle/\langle k\rangle-1)$, where $\langle k\rangle$ denotes the average degree of the network and $\langle k^2\rangle$ is the second moment of the degree distribution. This work was later generalized in Refs.~\cite{newman2002assortative,newman2003mixing,goltsev2008percolation} to study node percolation in networks with degree-degree correlations, which we will briefly review below. 

Consider a random network where $P(k)$ is the probability that a node has $k$ connections (with $1\leq k_{\min}\leq k \leq k_{\max}$).  The excess degree distribution $P_1(\widetilde{k})=(\widetilde{k}+1)P(\widetilde{k}+1)/\langle k \rangle$ gives the probability of selecting a node (through an edge) with $\widetilde{k}=k-1$ outgoing connections. We further define $P_1(\widetilde{k},\widetilde{k'})$ as the joint probability that a randomly selected edge connects two nodes with outgoing degrees $\widetilde{k}$ and $\widetilde{k'}$. For the typical uncorrelated case, the degrees at the two ends of an edge are independent, and therefore $P_1(\widetilde{k},\widetilde{k'}) = P_1(\widetilde{k})P_1(\widetilde{k'})$. 

For a node percolation process on random networks with an infinite number of nodes, the relative size of the GC can be obtained by first solving the following system of self-consistent equations,
\begin{eqnarray}
f_{\infty,\widetilde{k}}&=&p\sum_{\widetilde{k'}} P_1(\widetilde{k'}|\widetilde{k})\left(1-(1-f_{\infty,\widetilde{k'}})^{\widetilde{k'}}\right),
\end{eqnarray}
with $\widetilde{k}\in[k_{\min}-1,k_{\max}-1]$. Here, $f_{\infty,\widetilde{k}}$ denotes the probability that a node with $\widetilde{k}$ outgoing links is connected to the GC, and $P_1(\widetilde{k}|\widetilde{k'})$ is the conditional probability $P_1(\widetilde{k}|\widetilde{k'}) = P_1(\widetilde{k},\widetilde{k'})/\sum_{\widetilde{k}}P_1(\widetilde{k},\widetilde{k'})$.  With this system solved, the relative size of the GC is given by,
\begin{eqnarray}\label{eq.Pinfcorr}
P_{\infty}=p\left(1-\sum_k P(k) (1-f_{\infty,k-1})^k\right).
\end{eqnarray}
On the other hand, the critical point $p_c$ is determined by the condition that the largest eigenvalue $\lambda$ of the branching matrix is equal to 1. That is, by setting $p=p_c$, the critical threshold is obtained from the equation,
\begin{eqnarray}\label{eq.Branchcorr}
    \lambda(B)=1,
\end{eqnarray}
where the branching matrix $B$ is defined as,
\begin{eqnarray}
B_{\widetilde{k},\widetilde{k'}}=p\left(\widetilde{k}P_1(\widetilde{k}|\widetilde{k'})\right).
\end{eqnarray}
Additional discussion of percolation in uncorrelated and correlated networks can be found in Refs.~\cite{callaway2000network,newman2002assortative,newman2003mixing,goltsev2008percolation,vazquez2003resilience}.

\section{Percolation in Correlated Dyadic Networks}\label{sec.PerCorrDy}
In this section, we present our results on node percolation in correlated networks with dyadic interactions.

As mentioned in the Introduction, several studies have explored random node percolation in networks with assortative and disassortative mixing. In order to study the effect of degree correlations, a common approach is to keep the degree distribution $P(k)$ fixed while varying the joint degree distribution $P_1(\widetilde{k},\widetilde{k'})$, so that any observed effects can be attributed to degree correlations only. This same strategy of isolating correlations from degree effects has also been used in clustered configuration model networks,  where the inter-subgraph mixing is varied while the joint degree distribution and the clustering coefficient are held fixed~\cite{mann2025mixing}.

Here, we will show that for some networks with the same degree distribution $P(k)$, increasing disassortativity can actually make the network more robust. We will illustrate this behavior using trimodal networks, which we describe below, and then we will study the case of correlated networks whose degree distribution follows a truncated Poisson and power-law function.

\subsection{Degree-Degree Correlation in Trimodal Networks}
\subsubsection{Mathematical formulation}
We consider a network in which nodes can take only three possible degrees: $k_1$, $k_2$, or $k_3$, where $1\leq k_1<k_2<k_3$. Such a network will be referred to as a trimodal network, and its degree distribution is given by:
\begin{eqnarray}
P(k) = P(k_1)\delta_{k,k_1}+P(k_2)\delta_{k,k_2}+P(k_3)\delta_{k,k_3},
\end{eqnarray}
where $\delta$ is the Kronecker delta. From this distribution, the excess-degree distribution follows as
\begin{eqnarray} \label{eq.P1pi}
P_1(\widetilde{k_i})=\frac{(\widetilde{k_i}+1)P(\widetilde{k_i}+1)}{\langle k\rangle}\equiv \pi_i,
\end{eqnarray}
with,
\begin{eqnarray}
    \langle k\rangle =k_1P(k_1)+k_2P(k_2)+k_3P(k_3),
\end{eqnarray}
and $\widetilde{k_i}=k_i-1$ for $i=1,2,3$.

Since the present trimodal network contains no isolated nodes (as $k_1 \geq 1$),  there exists a one-to-one correspondence between the excess-degree distribution $P_1(\widetilde{k_i})$ and the original degree distribution $P(k_i)$ through the inverse relation $P(k_i)=\pi_i\langle k\rangle/k_i$ [see Eq.~(\ref{eq.P1pi})]. As a consequence, fixing the excess-degree distribution $P_1(\widetilde{k_i})=\pi_i$ automatically fixes the degree distribution $P(k_i)$. This differs from the general case discussed in Ref.~\cite{newman2002assortative}, where the presence of degree-zero nodes prevents the complete reconstruction of $P(k)$ from the excess-degree distribution alone.

We will now introduce the joint excess-degree distribution $P_1(\widetilde{k},\widetilde{k'})$. For a trimodal network, this probability can be written as,
\begin{eqnarray}
P_1(\widetilde{k},\widetilde{k'})  =\sum_{i=1}^{3}\sum_{j=1}^{3}P_1(\widetilde{k_i},\widetilde{k_j})\delta_{\widetilde{k},\widetilde{k_i}}\delta_{\widetilde{k'},\widetilde{k_j}},
\end{eqnarray}
This joint probability can be conveniently represented as a $3\times 3$ matrix:
\begin{equation}
P_1(\widetilde{k},\widetilde{k'})=\begin{pmatrix}
P_1(\widetilde{k_1},\widetilde{k_1}) & P_1(\widetilde{k_1},\widetilde{k_2}) & P_1(\widetilde{k_1},\widetilde{k_3}) \\
P_1(\widetilde{k_2},\widetilde{k_1}) & P_1(\widetilde{k_2},\widetilde{k_2}) & P_1(\widetilde{k_2},\widetilde{k_3}) \\
P_1(\widetilde{k_3},\widetilde{k_1}) & P_1(\widetilde{k_3},\widetilde{k_2}) & P_1(\widetilde{k_3},\widetilde{k_3})
\end{pmatrix}
\end{equation}
We assume this matrix to be symmetric and impose the constraint that the sum over each row (or column) yields the corresponding excess-degree probability $\pi_i$:
\begin{eqnarray}
\sum_{j=1}^{3} P_1(\widetilde{k_i},\widetilde{k_j})=\pi_i. 
\end{eqnarray}
The Pearson assortativity coefficient $r$ can be computed from the joint excess-degree distribution as:
\begin{eqnarray}\label{eq.corrGen}
r = \frac{\sum_{\widetilde{k},\widetilde{k'}} \widetilde{k}\,\widetilde{k'} \left[P_1(\widetilde{k},\widetilde{k'}) - P_1(\widetilde{k})P_1(\widetilde{k'})\right]}{\sigma_{\widetilde{k}}^2},
\end{eqnarray}
where $\sigma_{\widetilde{k}}^2 = \sum_{\widetilde{k}} \widetilde{k}^2 P_1(\widetilde{k}) - \left[\sum_{\widetilde{k}} \widetilde{k} P_1(\widetilde{k})\right]^2$ is the variance of the excess-degree distribution.

In order to explore networks with different degree–degree correlations while keeping the degree distribution fixed, we now introduce the following parametrization:
\begin{equation}\label{eq.Pkkalphas}
P_1(\widetilde{k},\widetilde{k'})=\begin{pmatrix}
\pi_1-\alpha_1-\alpha_2& \alpha_1 & \alpha_2 \\
\alpha_1 & \pi_2-2\alpha_1 & \alpha_1 \\
\alpha_2 & \alpha_1 & \pi_3-\alpha_1-\alpha_2
\end{pmatrix},
\end{equation}
where $\pi_1$, $\pi_2$ and $\pi_3$ are known and $\alpha_1$ and $\alpha_2$ are two free parameters subject to positivity and normalization constraints. By construction, this parametrization ensures that the matrix $P_1(\widetilde{k},\widetilde{k'})$ is symmetric and preserves the marginal distributions $P_1(\widetilde{k_i})=\pi_i$, which is independent of $\alpha_1$ and $\alpha_2$. Consequently, the excess-degree distribution (and therefore the degree distribution $P(k)$) remains fixed as we vary $\alpha_1$ and $\alpha_2$, which will allow us to systematically study networks with different degree-degree correlations while keeping the degree distribution constant. This construction, which preserves the marginals, is similar in spirit to the mixing-matrix approach used by Mann et al.~\cite{mann2025mixing} for inter-subgraph correlations in clustered configuration model networks.

For the particular parametrization given in Eq.~(\ref{eq.Pkkalphas}), the Pearson correlation coefficient is given by [after some algebra using Eqs.~(\ref{eq.corrGen}) and~(\ref{eq.Pkkalphas})]:
\begin{eqnarray}\label{eq.rtrimod}
r=1-\alpha_1\left(\frac{(k_1-k_2)^2+(k_2-k_3)^2}{\frac{\langle k(k-1)^2\rangle}{\langle k\rangle}-\left(\frac{\langle k(k-1)\rangle}{\langle k\rangle}\right)^2}\right)-\alpha_2\left(\frac{(k_1-k_3)^2}{\frac{\langle k(k-1)^2\rangle}{\langle k\rangle}-\left(\frac{\langle k(k-1)\rangle}{\langle k\rangle}\right)^2}\right).
\end{eqnarray}

\subsubsection{Numerical results}
\begin{figure}
  \subfigure[]{
    \begin{minipage}{.48\columnwidth}
      \centering
      \includegraphics[scale=0.40]{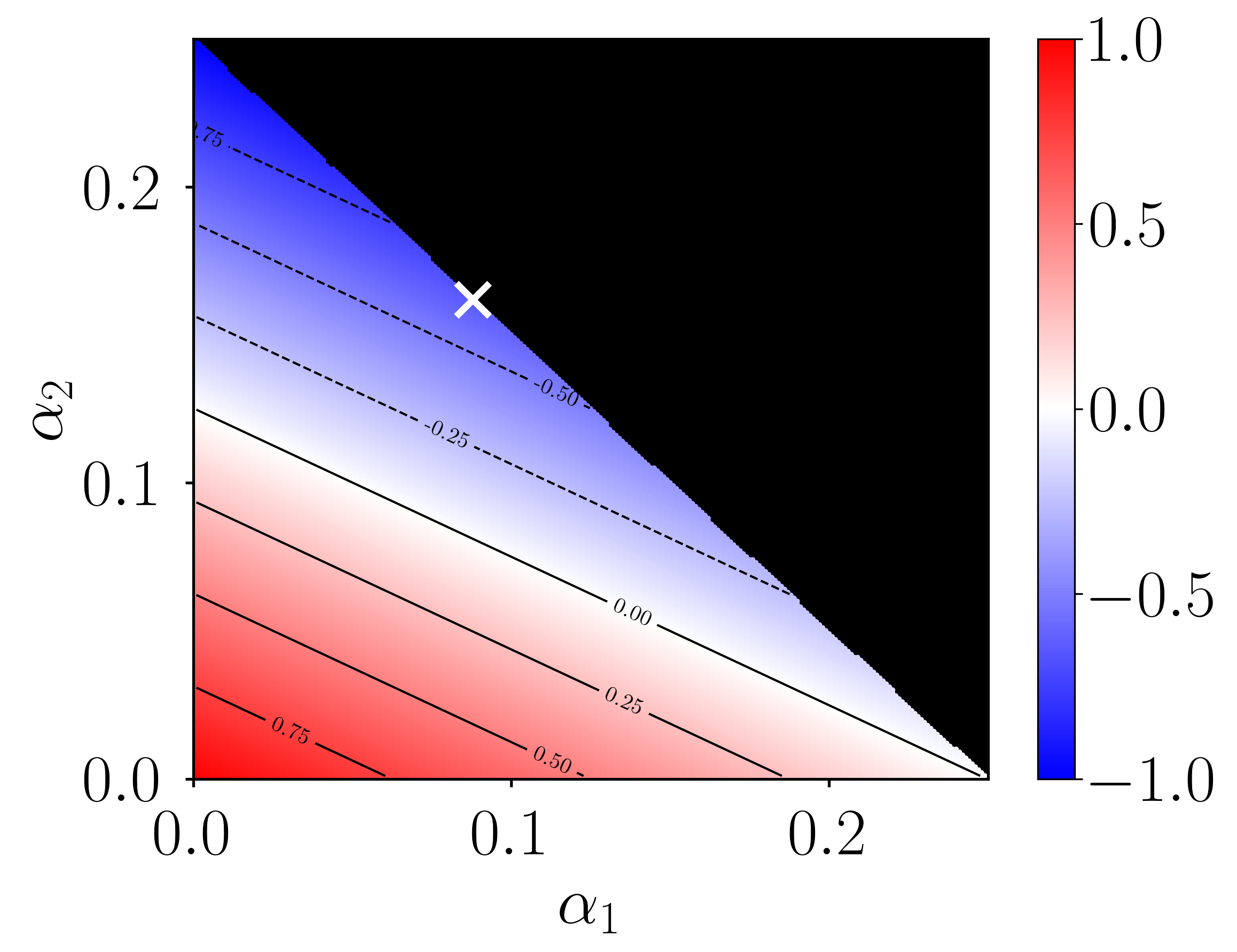}
    \end{minipage}
  }\hfill
  \subfigure[]{
    \begin{minipage}{.48\columnwidth}
      \centering
      \includegraphics[scale=0.30]{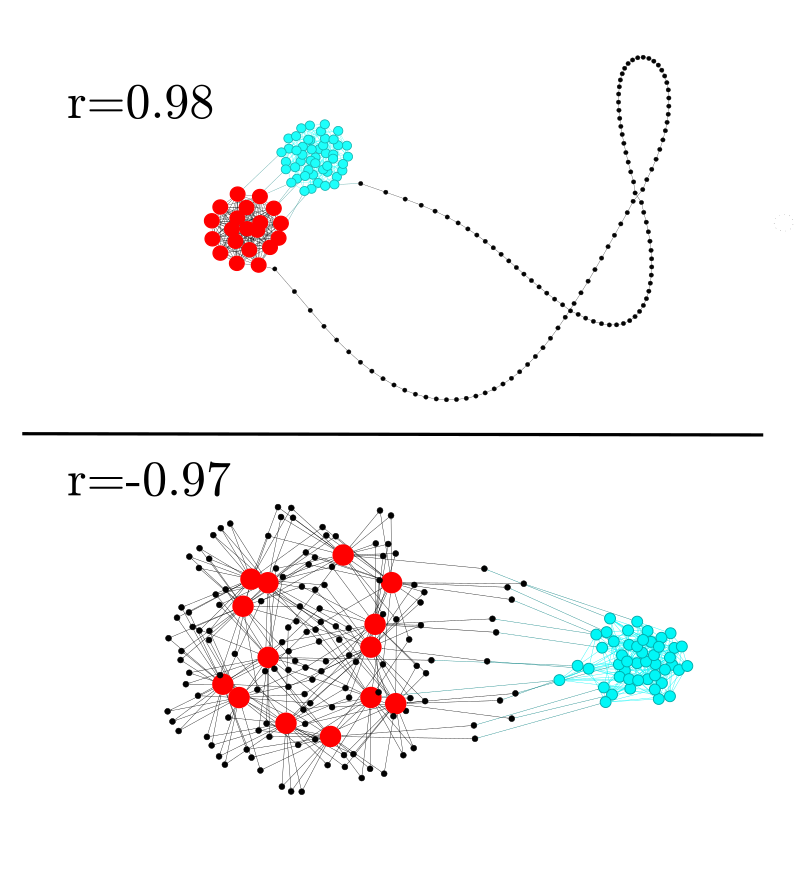}
    \end{minipage}
  }
  \caption{Panel (a): Heatmap of the Pearson assortativity coefficient $r$ in the $\alpha_1$-$\alpha_2$ plane for the trimodal network with correlation matrix given by Eq. (\ref{Eq.corrmatrix}). Straight lines correspond to constant values of $r$. The white cross marks the point of maximum fragility ($\alpha_1=0.088$, $\alpha_2=0.162$, $r=-0.65$), whose critical threshold is shown in Fig.~\ref{fig.trimPc}b.  Panel (b): Network visualizations generated using Gephi for two different correlation values: $r=-0.97$ and $r=0.98$. Networks were generated through numerical simulations with $N=1000$ nodes. Light blue circles correspond to nodes with intermediate degree $k_2=10$, small black circles represent nodes with low degree $k_1=2$, and large red circles represent nodes with degree $k_3=19$. In the highly assortative case ($r=0.98$), three well-defined communities emerge, each composed of nodes with the same degree. In the highly disassortative case ($r=-0.97$), two communities are observed: one formed exclusively by intermediate-degree nodes, and another with a bipartite structure connecting nodes with low and high degree.}
  \label{fig.corrPlane}
\end{figure}

Consider a trimodal network where $k_1=2$, $k_2=10$, $k_3=19$, and $\pi_1=0.25$, $\pi_2=0.50$, $\pi_3=0.25$. The degree distribution $P(k_i)$ is obtained from the probabilities $\pi_i$ through the relation $P(k_i)=\langle k\rangle \pi_i/k_i$, leading to $P(k_1)=0.66$, $P(k_2)=0.27$, and $P(k_3)=0.07$. On the other hand, for this network, the joint probability matrix becomes:

\begin{equation}
P_1(\widetilde{k},\widetilde{k'})=\begin{pmatrix}
0.25-\alpha_1-\alpha_2& \alpha_1 & \alpha_2 \\
\alpha_1 & 0.50-2\alpha_1 & \alpha_1 \\
\alpha_2 & \alpha_1 & 0.25-\alpha_1-\alpha_2
\end{pmatrix}
\label{Eq.corrmatrix}
\end{equation}
In Fig.~\ref{fig.corrPlane}a, we show the Pearson assortativity coefficient [computed using Eq.~(\ref{eq.rtrimod})] in the $\alpha_1$-$\alpha_2$ plane for the joint probability matrix given above. From this figure, we can observe that curves of constant correlation are straight lines.  This is a direct consequence of Eq.~(\ref{eq.rtrimod}), which establishes a linear dependence between $\alpha_1$ and $\alpha_2$ when $r$ is held constant. As the figure illustrates, this parametrization allows us to model a wide range of degree correlations $r$. The extreme positive correlation, $r=1$, is achieved when $\alpha_1=\alpha_2=0$, which corresponds to an entirely diagonal correlation matrix $P_1(\widetilde{k},\widetilde{k'})$, meaning that in this scenario, nodes connect exclusively with nodes of equal degree.  An example of a highly assortative network is displayed in Fig.~\ref{fig.corrPlane}b, where we can see that three clearly defined communities emerge (each associated with nodes of degree $k_i$), which is in agreement with the well-known connection between assortativity and modular structure explored in detail in Ref.~\cite{van2010spectral}. On the other hand, the most negative correlation accessible to this trimodal network is $r\approx-0.99$, corresponding to $\alpha_1=0$ and $\alpha_2=0.25$. Similar to the assortative case, in Fig.~\ref{fig.corrPlane}b we show an example of a network with correlation $r=-0.98$, where we observe two groups or communities. One is composed exclusively of nodes with intermediate degree $k_2=10$, which can be regarded as a random regular graph of degree 10. The other has a bipartite structure, and connects nodes of connectivity $k_1=2$ with $k_3=19$ (i.e., those with the lowest and highest degree). These results are in agreement with those previously found by Jing et al~\cite{jing2007effects}.

Since the trimodal network can span nearly the full possible correlation range $[-1,1]$, we will now explore in detail how different levels of correlation affect network robustness.

\begin{figure}[ht]
  \subfigure[]{
    \begin{minipage}{.48\columnwidth}
      \centering
      \includegraphics[scale=0.48]{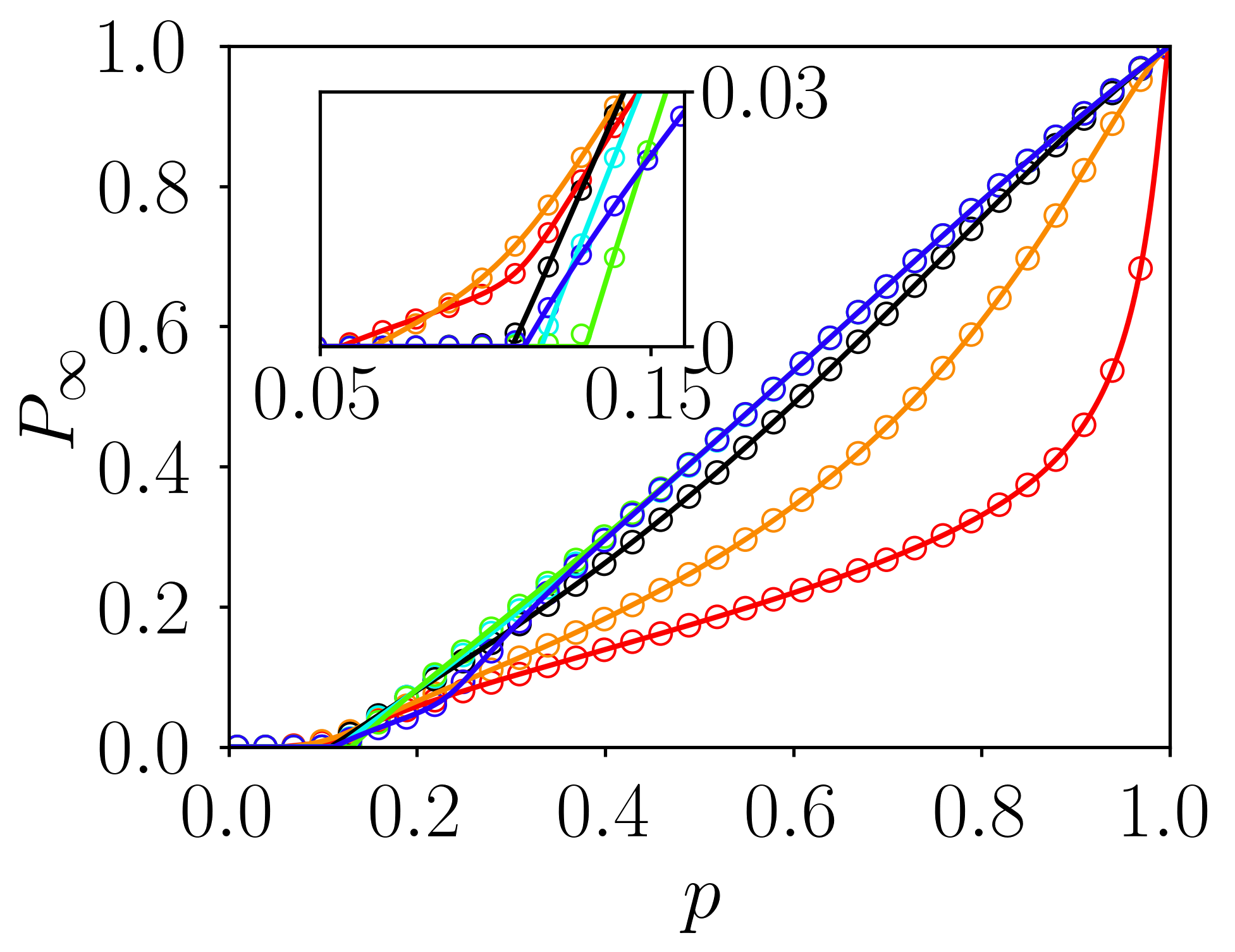}
    \end{minipage}
  }\hfill
  \subfigure[]{
    \begin{minipage}{.48\columnwidth}
      \centering
      \includegraphics[scale=0.40]{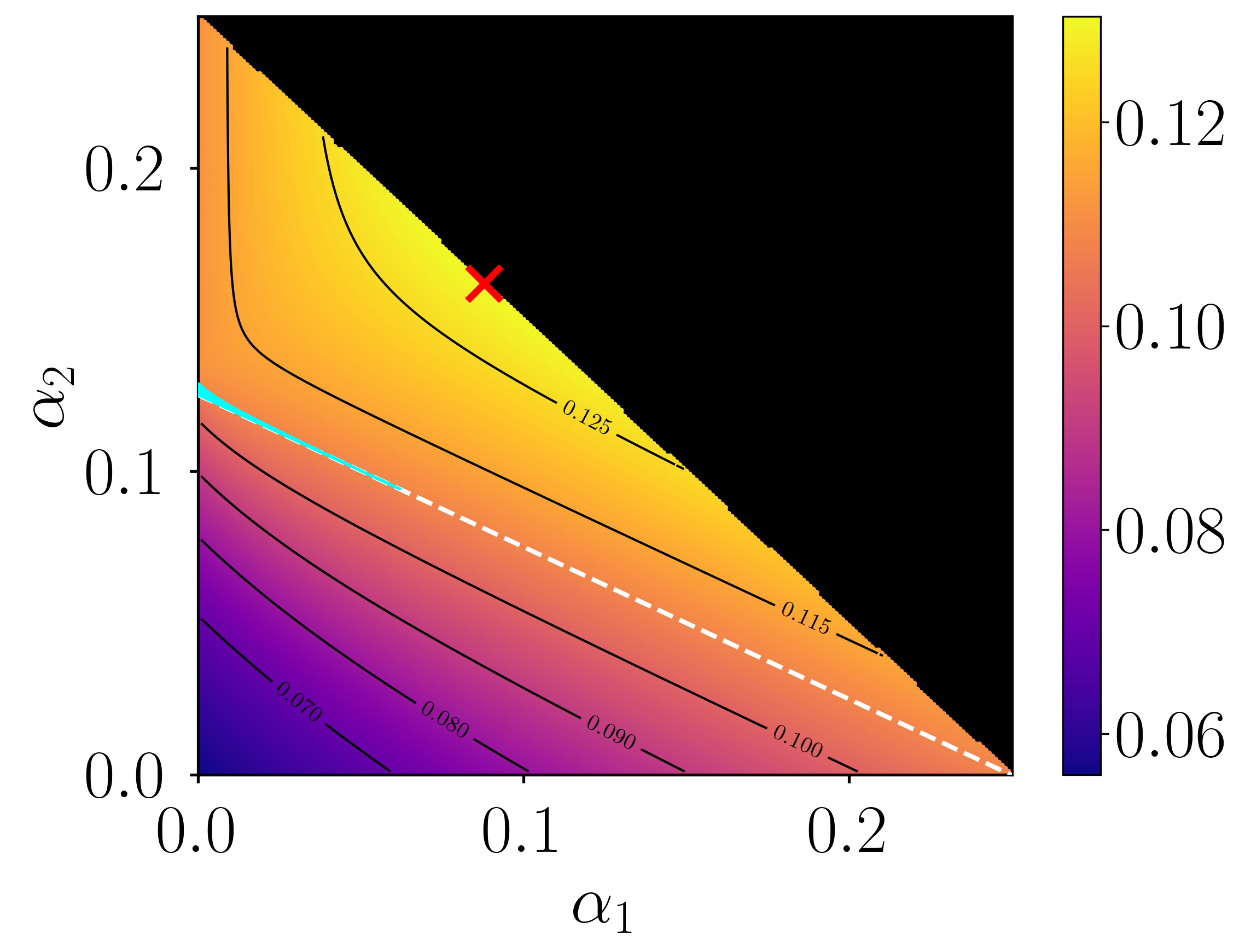}
    \end{minipage}
  }
\caption{Panel (a): Relative size of the giant component, $P_{\infty}$, as a function of $p$ for several correlation values. The curves correspond to $r=0.98$ ($\alpha_1=0.003$, $\alpha_2=0.001$, red), $r=0.75$ ($\alpha_1=0.02$, $\alpha_2=0.0213$, orange), $r=0$ ($\alpha_1=0.0503$, $\alpha_2=0.10$, black), $r=-0.2$ ($\alpha_1=0.2001$, $\alpha_2=0.0498$, light blue), $r=-0.65$ ($\alpha_1=0.088$, $\alpha_2=0.162$, green), and $r=-0.99$ ($\alpha_1=0.0016$, $\alpha_2=0.2484$, blue). Solid lines correspond to theoretical predictions from Eq.~(\ref{eq.Pinfcorr}), while symbols represent numerical simulations on networks with $N=10^6$ nodes averaged over 100 stochastic realizations. The inset shows a magnification of the main plot near the transition point. Panel (b): Heatmap of the critical percolation threshold $p_c$ in the $\alpha_1$-$\alpha_2$ plane predicted by Eq.~(\ref{eq.Branchcorr}). Near the origin ($\alpha_1=\alpha_2=0$), $p_c$ approaches $1/(k_3-1)=1/18\approx 0.0555$ (for example, $p_c=0.056$ for $\alpha_1=0.001$, $\alpha_2=0.001$), which indicates that the high-degree subnetwork sustains the GC in strongly assortative networks. Near the point $(\alpha_1,\alpha_2)=(0,0.25)$, $p_c$ approaches $1/(k_2-1)=1/9\approx 0.1111$ (for example, $p_c=0.1115$ for $\alpha_1=0.001$, $\alpha_2=0.249$), showing that the intermediate-degree community controls robustness in highly disassortative networks. The red cross marks the point of maximum fragility ($\alpha_1=0.088$, $\alpha_2=0.162$, $r=-0.65$).}
\label{fig.trimPc}
\end{figure}

In Fig.~\ref{fig.trimPc}a we plot $P_{\infty}$ as a function of $p$ for several values of $r$. As clearly shown in the inset, positively correlated networks display the smallest values of $p_c$, as expected, and as $r$ decreases from $r=1$, the critical threshold increases, meaning that the system becomes more fragile. However, $p_c$ exhibits a nonmonotonic behavior as $r$ approaches $r=-1$.

To fully visualize how the value of $p_c$ depends on the parameters $\alpha_1$ and $\alpha_2$, in Fig.~\ref{fig.trimPc}b we show $p_c$ in the $\alpha_1$-$\alpha_2$ plane. First, motivated by the work of Wang et al.~\cite{wang2022basic}, we explore how the percolation threshold $p_c$ of correlated networks compares with the value $p_c = 1/(\langle k^2\rangle/\langle k\rangle - 1)$ corresponding to an uncorrelated network generated by the configuration model (see Sec.~\ref{sec.Backg}). In Fig.~\ref{fig.trimPc}b, we highlight in light blue the region where disassortative networks are more robust than the fully uncorrelated case. This result, in agreement with Wang et al.~\cite{wang2022basic}, indicates that the relation between degree correlations and $p_c$ is not trivial. It is worth emphasizing that this behavior is observed close to the region of the plane where $r=0$. However, as we will show below, our model also exhibits nonmonotonic robustness in the strongly disassortative regime.

Another point that emerges from Fig.~\ref{fig.trimPc}b is that near the origin (that is, $\alpha_1=\alpha_2=0$) the isolines of constant $p_c$ are approximately linear, which corresponds to the region where networks are assortative. Furthermore, as we approach the origin, the value of $p_c$ decreases and approaches $p_c=1/(19-1)$, which corresponds to the critical point of a random regular network with connectivity $k=19$. This shows that in strongly assortative configurations, the subnetwork or community formed by the highest-degree nodes constitutes the core sustaining the GC.

On the other hand, from Fig.~\ref{fig.trimPc}b, near the upper boundary $\alpha_2=0.25-\alpha_1$,  the isolines bend sharply and become almost vertical close to the point $(\alpha_1,\alpha_2)=(0,0.25)$. Consequently, under strongly disassortative mixing, $p_c$ becomes mainly sensitive to $\alpha_1$, which is the parameter controlling how nodes of intermediate degree $k_2=10$ are connected [see second row in the matrix given in Eq.~(\ref{Eq.corrmatrix})]. Analogous to the assortative case, we obtain that in the limit $r\to -0.99$, the value of $p_c$ approaches $p_c=1/(10-1)$, which corresponds to the critical point of a random regular network with connectivity $k=10$.  Therefore, in highly disassortative networks, the intermediate-degree community is the one that sustains the GC and sets the value of $p_c$.

Finally, in Fig.~\ref{fig.trimPc}b, we mark the point of maximum fragility (highest value of $p_c$), which corresponds to the point with coordinates ($\alpha_1\approx 0.088$, $\alpha_2\approx 0.162$), which has a correlation of $r=-0.65$. Our results, therefore indicate that although disassortative networks are more fragile than assortative ones, the most fragile configuration does not coincide with the maximally negative Pearson coefficient. Similar results for other trimodal dyadic networks are presented in Appendix~\ref{appDyaNet}.

\begin{figure}[H]
  \centering
  \includegraphics[scale=0.70]{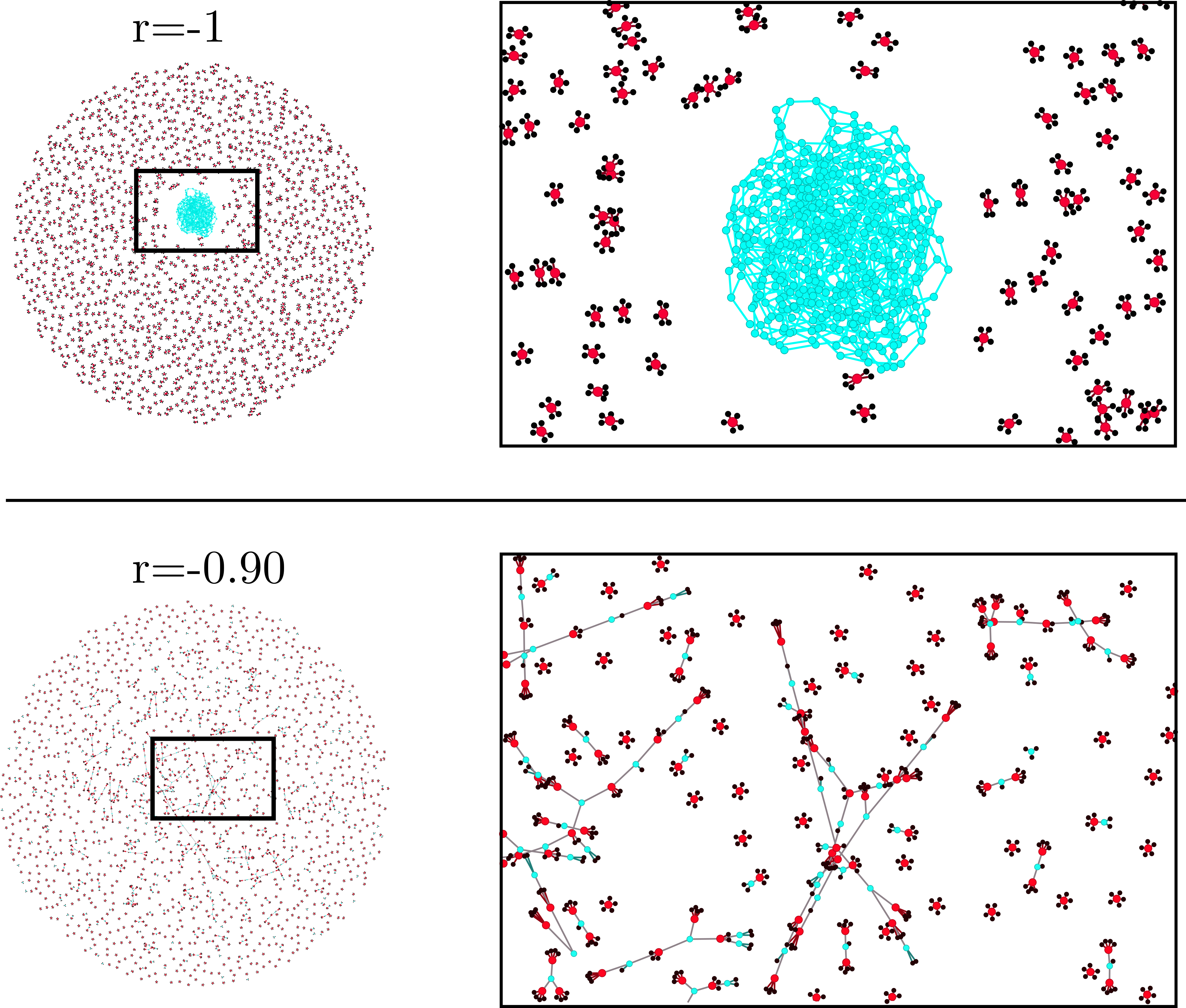}
\caption{Snapshots of networks illustrating that networks with $r=-1$ are not always the most fragile configuration.  Both networks have $N=1900$ nodes and the same degree distribution  ($k_1=1$, $k_2=3$, $k_3=5$) with  $P(k_1)=15/19$, $P(k_2)=1/19$, and $P(k_3)=3/19$.  Nodes are colored according to their degree: black ($k=1$), light blue ($k=3$), and red ($k=5$).  Top panel: maximally disassortative configuration ($r=-1$). A zoomed-in view of the area within the black rectangle is provided on the right. High-degree nodes ($k=5$) connect exclusively to low-degree nodes ($k=1$), forming many small clusters, while intermediate-degree nodes ($k=3$) connect only among themselves, generating a GC.  Bottom panel: slightly weaker disassortative mixing ($r\approx -0.90$).  Intermediate-degree nodes now connect to both high- and low-degree nodes, which disrupts the connected $k_2=3$ subnetwork.  As a result, the GC disappears, and the network decomposes into small tree-like clusters.  Network visualizations were generated using Gephi~\cite{bastian2009gephi}.}
\label{fig.Esquemat}
\end{figure}

To gain intuition about why networks with $r \approx -1$ are less fragile than moderately disassortative networks, consider a trimodal network with $k_1=1$, $k_2=3$, and $k_3=5$, and $P(k_1)=15/19$, $P(k_2)=1/19$, and $P(k_3)=3/19$. We also assume that the joint distribution $P_1(\widetilde{k},\widetilde{k'})$ follows the form given in Eq.~(\ref{Eq.corrmatrix}). For $\alpha_1=0$ and $\alpha_2=5/11\approx 0.4545$, the network has a maximally negative correlation $r=-1$, which is illustrated in the upper panel of Fig.~\ref{fig.Esquemat}. As we can see, nodes with the highest degree ($k=5$) connect exclusively with the lowest-degree nodes ($k=1$), forming a large number of small clusters. At the same time, the network possesses a GC formed exclusively by intermediate-degree nodes ($k=3$), so the value of $p_c$ corresponds to the critical threshold of a random regular network of degree 3, namely $p_c = 1/(3-1) = 1/2$.

We now consider a slightly less disassortative network with the same degree distribution but correlation $r\approx -0.90$, corresponding to $\alpha_1=0.0454$ and $\alpha_2=0.409$ (lower panel of Fig.~\ref{fig.Esquemat}). 
In this configuration, intermediate-degree nodes no longer connect exclusively among themselves, but instead connect to both high- and low-degree nodes, which makes the network less disassortative. In addition, we also observe that intermediate-degree nodes no longer link to each other, and as a result, the GC that existed at $r=-1$ disappears entirely. One might expect that, as we move from $r=-1$ to $r=-0.90$, the newly formed connections between high- and intermediate-degree nodes would create a denser core, making the network more robust. However, the opposite occurs. This can be verified theoretically from Eq.~(\ref{eq.Branchcorr}), which yields $p_c = 1.6 > 1$, indicating that no GC exists for $r=-0.90$ even before any nodes are removed. This is also apparent from the figure, where clusters are small and tree-like.

In other words, although intermediate-degree nodes start connecting to high-degree nodes when the correlation increases from $r=-1$ to $r\approx -0.90$, this does not lead to the formation of a robust core. Instead, these new connections destroy the GC of intermediate-degree nodes. Moreover, the high-degree nodes remain predominantly attached to many low-degree nodes ($k=1$), which prevents them from forming a new GC, or at least a sufficiently connected backbone. The network therefore becomes more fragile, as the correlation increases from $r=-1$.

\subsection{Percolation in Networks with Poisson and Power-Law Degree Distributions}\label{sec.dyadCop}

In this section, we present our results for node percolation on random networks whose degree distribution follows either a Poisson or a power-law form. Similar to the trimodal network case, we will modify the joint excess-degree distribution $P_1(\widetilde{k},\widetilde{k'})$ while keeping the excess-degree distribution $P_1(\widetilde{k})$ fixed, so that any observed effects on network robustness can be attributed to degree-degree correlations only. As mentioned previously, if $P_1(\widetilde{k})$ is fixed
and the network doesn't contain nodes with degree zero (i.e., $k_{\min} \geq 1$), then the mapping between $P_1(\widetilde{k})$ and $P(k)$ is invertible, and therefore the degree distribution  $P(k)$ is also fixed.

Specifically, we will study networks with a truncated Poisson distribution given by:
\begin{equation}
\text{Pois}(\gamma,k_{\min},k_{\max}) = 
\begin{cases} 
      c \frac{\gamma^k\exp(-\gamma)}{k!}, & \text{if } k_{\min}\leq k \leq k_{\max} \\
      0, & \text{otherwise}
\end{cases}
\label{eq.trunPois}
\end{equation}
where $c$ is a normalization constant, and networks with a truncated power-law distribution $\text{PL}(\gamma,k_{\min},k_{\max})$, defined as:
\begin{equation}
     \text{PL}(\gamma,k_{\min},k_{\max}) = 
     \begin{cases} 
          c k^{-\gamma}, & \text{if } k_{\min}\leq k \leq k_{\max} \\
          0, & \text{otherwise}
     \end{cases}
\label{eq.trunPL}
\end{equation}
where $c$ is also a normalization constant.

For trimodal networks, it was relatively easy to find a parametrization of $P_1(\widetilde{k},\widetilde{k'})$ compatible with a given excess-degree distribution. This parametrization, presented in Eq.~(\ref{Eq.corrmatrix}), depends on two variables, $\alpha_1$ and $\alpha_2$, and allowed us to explore a wide range of correlation values $r$ while keeping $P_1(\widetilde{k})$ fixed. However, for networks with more general degree distributions, finding a compatible parametrization is more challenging, as it would require introducing a larger number of parameters $\alpha_i$. Therefore, we will employ a different approach based on the Iterative Proportional Fitting (IPF) algorithm~\cite{lomax2016estimating,norman1999putting}.

The IPF algorithm has been widely used in statistics and, in general, allows one to estimate a bivariate probability distribution $p(x,y)$ given the following marginal constraints: $\sum_y p(x,y)=p(x)$ and $\sum_x p(x,y)=p(y)$, where $x$ and $y$ are discrete variables. As explained in Ref.~\cite{norman1999putting}, the algorithm proceeds as follows:
\begin{itemize}
    \item Step 1: An initial seed matrix $p^{(0)}(x,y)$ is proposed.
    \item Step 2: The rows of the matrix are rescaled so that the row sums match the target marginal distribution $p(x)$.
    \item Step 3: The columns of the matrix are rescaled so that the column sums match the target marginal distribution $p(y)$.
\end{itemize}
Steps 2 and 3 are repeated iteratively until the error between the current marginals and the target marginals falls below a specified threshold.

In order to introduce a tunable level of correlation in a controlled manner,  we will use a copula-based construction to create the initial seed matrix. In particular, we employ the Clayton copula~\cite{zhang2020quantification} defined as,
\begin{equation}
C_{\theta}(u,v) = \max\{(u^{-\theta} + v^{-\theta} - 1)^{-1/\theta},0\} \quad \text{with } \theta \in [-1,\infty)\setminus\{0\},
\end{equation}
which has the advantage of depending on a single parameter $\theta$ and will allow us to smoothly interpolate between assortative and disassortative structures. This choice is convenient but not unique and other copulas could be used as well.

To implement this construction, let $F(\widetilde{k})$ denote the cumulative distribution function associated with the excess-degree distribution $P_1(\widetilde{k})$, defined as
\begin{equation}
F(\widetilde{k}) = \sum_{\widetilde{j} \le \widetilde{k}} P_1(\widetilde{j}).
\end{equation}
We then construct the initial seed matrix $P_1^{(0)}(\widetilde{k},\widetilde{k'})$ using a standard copula-based discretization~\cite{geenens2020copula,raschke2014copula} as follows:
\begin{eqnarray}
P_1^{(0)}(\widetilde{k},\widetilde{k'}) &=& C_{\theta}\left( F(\widetilde{k}),F(\widetilde{k'})\right)-C_{\theta}\left( F(\widetilde{k}-1),F(\widetilde{k'})\right)-C_{\theta}\left( F(\widetilde{k}),F(\widetilde{k'}-1)\right)+\nonumber\\
&&+C_{\theta}\left( F(\widetilde{k}-1),F(\widetilde{k'}-1)\right).
\end{eqnarray}

After defining the seed matrix, Steps 2 and 3 of the IPF algorithm are applied iteratively until the prescribed marginal distributions $P_1(\widetilde{k})$ are recovered within a tolerance of $10^{-10}$.

Once the final matrix $P_1(\widetilde{k},\widetilde{k'})$ is obtained through this procedure, we solve the equations presented in the previous sections in order to compute $P_{\infty}(p)$ and $p_c$.

\begin{figure}[ht]
 \subfigure[]{
    \begin{minipage}{.48\columnwidth}
      \centering
      \includegraphics[scale=0.45]{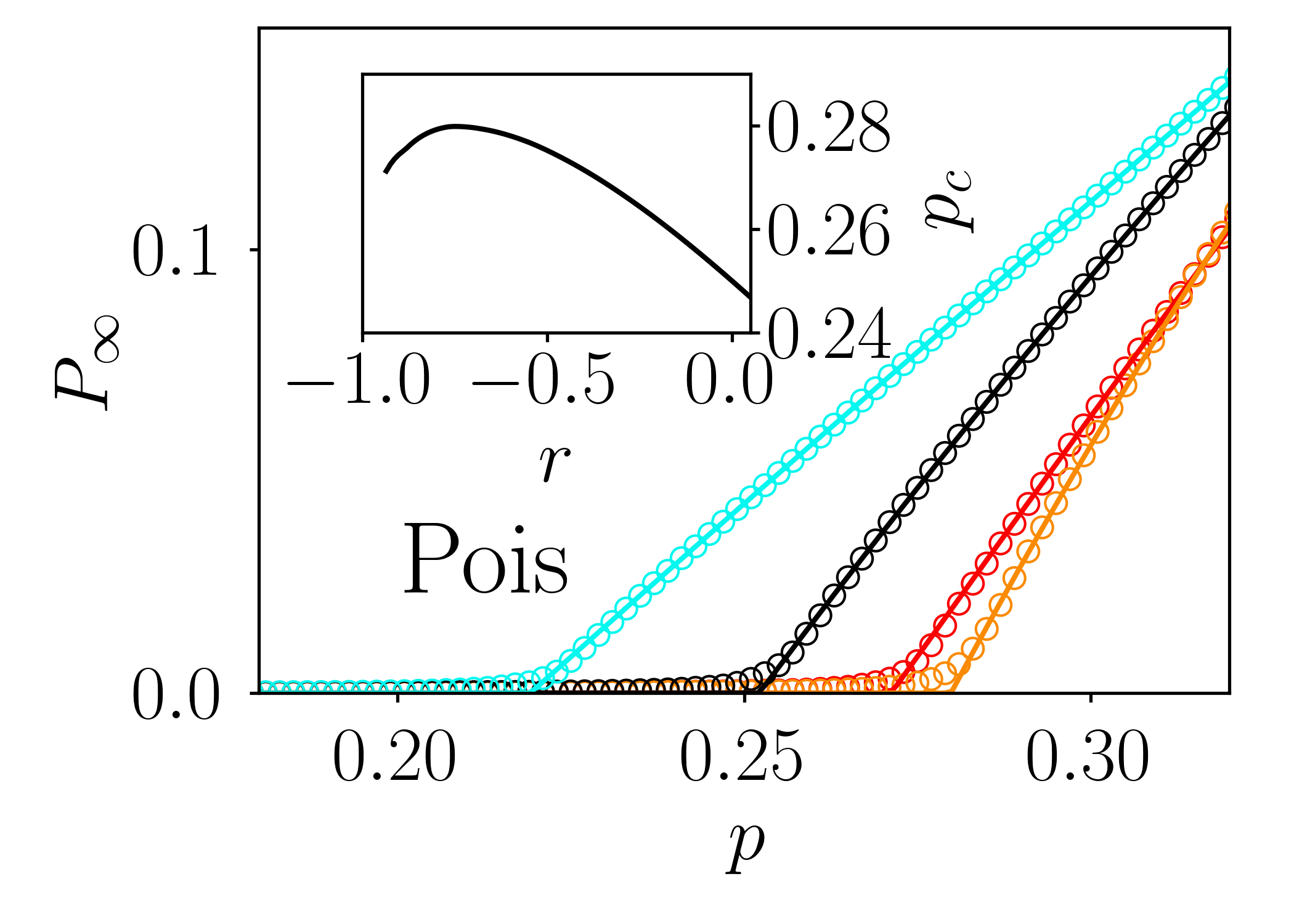}
    \end{minipage}
  }\hfill
  \subfigure[]{
    \begin{minipage}{.48\columnwidth}
      \centering
      \includegraphics[scale=0.45]{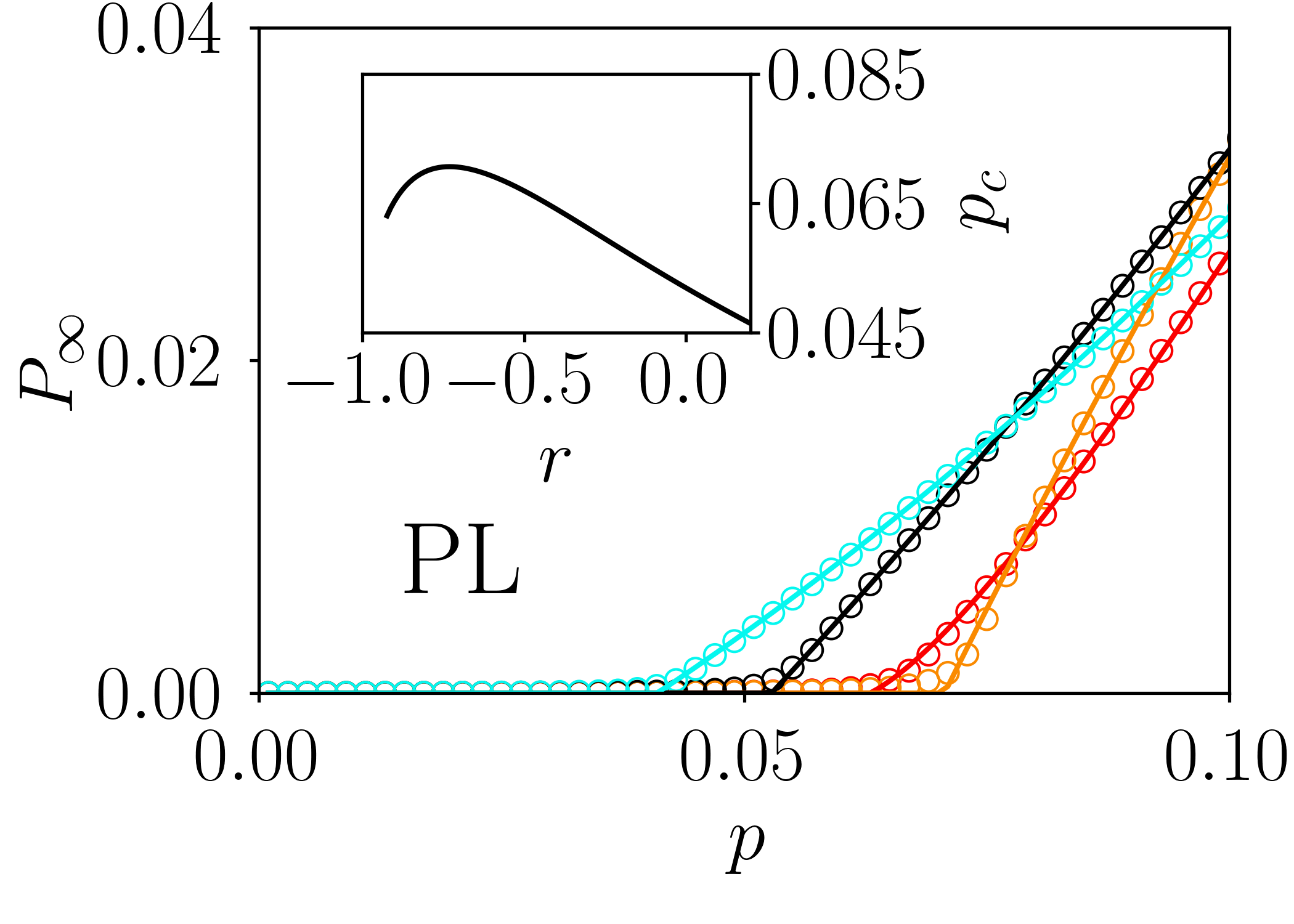}
    \end{minipage}
  }
\caption{Panel (a): In the main panel, we plot $P_{\infty}$ as a function of $p$ for networks with a truncated Poisson distribution  $\mathrm{Pois}(4,1,20)$. Solid lines correspond to the theoretical solution of Eq.~(\ref{eq.Pinfcorr}), symbols represent numerical simulations on networks with $N=10^6$ nodes averaged over 100 stochastic realizations. Colors indicate different values of the degree-degree correlation coefficient $r$: red $r=-0.93$ (obtained with $\theta=-0.99$ in the Clayton copula), orange $r=-0.75$ ($\theta=-0.81$), black $r=-0.04$ ($\theta=-0.05$), and light blue $r=0.43$ ($\theta=1.00$). The inset shows $p_c$ versus $r$ obtained from Eq.~(\ref{eq.Branchcorr}).
Panel (b): Same as in (a) but for a scale-free network with degree distribution $\mathrm{PL}(1.5,2,50)$. Colors indicate different values of the degree-degree correlation coefficient $r$: red $r=-0.92$ ($\theta=-0.99$), orange $r=-0.73$ ($\theta=-0.82$), black $r=-0.03$ ($\theta=-0.05$), and light blue $r=0.41$ ($\theta=1.00$).}
\label{fig.ERSF}
\end{figure}

In Fig.~\ref{fig.ERSF}a-b we present our results from the theory together with those obtained from stochastic simulations (see Appendix~\ref{app.Construct} for details on how correlated networks are generated in the simulations). As can be observed, both approaches reveal a nonmonotonic behavior as the degree–degree correlation is decreased, for both the truncated Poisson network and the scale-free network. The insets in the figures show more clearly the behavior of $p_c$ on $r$. We find that the maximum value of $p_c$ occurs for disassortative networks. However, as in the case of trimodal networks, this point of maximum fragility does not coincide with the limit of maximum disassortativity, showing that, in the thermodynamic limit, extreme negative correlations can have a non-trivial impact on the robustness of networks against random failures.

\section{Node Percolation in hypergraphs with Degree–Cardinality Correlations}\label{sec.HyperDegCar}

In this section, we will study how correlations affect the robustness of hypergraphs under a node percolation process. It is important to note that our goal is not to analyze every possible type of correlated hypergraph, but rather to provide a clear and concrete example showing that the relationship between the percolation threshold $p_c$ and the correlation coefficient $r$ can be, in some cases, nonmonotonic.

As mentioned in the Introduction, in a hypergraph, interactions do not occur exclusively between pairs of nodes, but can also involve groups of nodes. In dyadic networks, pairwise interactions are represented by links, whereas in hypergraphs they are described by hyperedges~\cite{battiston2021physics}. As discussed in several previous works, a hypergraph can be represented as a factor graph or, equivalently, as a bipartite network composed of two sets: one set contains the nodes of the original network, while the other contains factor nodes representing hyperedges. In this bipartite representation, a link between a node "$i$" and a factor node "$j$" indicates that node "$i$" belongs to group "$j$" (or, in other words, "$i$" is a member of "$j$"). Consequently, all nodes connected to the same factor node "$j$" interact collectively in the hypergraph.

The hyperdegree $k$ of a node is the number of groups it belongs to, or equivalently, the number of factor nodes it is connected to in the bipartite representation. The probability that a node has hyperdegree $k$ is denoted by $P(k)$. Analogously, the cardinality $m$ of a hyperedge is the number of nodes participating in that hyperedge, which corresponds in the bipartite representation to the number of nodes connected to a factor node. The probability that a hyperedge has cardinality $m$ is denoted by $Q(m)$.

Following the same approach used for dyadic networks, we define the excess-hyperdegree and 
excess-cardinality distributions in the bipartite representation as,
\begin{eqnarray}
P_1(\widetilde{k})&=&\frac{(\widetilde{k}+1)P(\widetilde{k}+1)}{\langle k\rangle},\\
Q_1(\widetilde{m})&=&\frac{(\widetilde{m}+1)Q(\widetilde{m}+1)}{\langle m\rangle},
\end{eqnarray}
where
$\langle k\rangle=\sum kP(k)$, and $\langle m\rangle=\sum mQ(m)$.

We now consider a percolation process in which a fraction $1-p$ of nodes is removed, along with all the hyperedges they belong to. In the bipartite representation, this implies that after removing a fraction $1-p$ of the nodes, any factor node that loses at least one of its neighbors is also removed. 

Before we generalize this process to correlated hypergraphs, it is useful to first review how it behaves in the simpler uncorrelated case, which was studied by Bianconi and Dorogovtsev~\cite{bianconi2024theory}. In that work, node hyperdegree and hyperedge cardinality are independent of each other, and the fraction of nodes belonging to the GC can be found by solving the following pair of self-consistent equations:
\begin{eqnarray}
W&=&\sum_{\widetilde{k}}P_{1}(\widetilde{k})\left[1-(1-V)^{\widetilde{k}}\right],\\
V&=&\sum_{\widetilde{m}}Q_{1}(\widetilde{m})p^{\widetilde{m}}\left[1-(1-W)^{\widetilde{m}}\right].
\end{eqnarray}
Here, $W$ denotes the probability that a node reached by following a randomly chosen link in the bipartite representation belongs to the GC, and $V$ is the probability that a factor node reached in the same way is connected to the GC.

Once we solve this system, the total fraction of nodes in the GC is:
\begin{eqnarray}
P_{\infty}=p\sum_k P(k)\left[1-(1-V)^k\right].
\end{eqnarray}
In addition, for the uncorrelated case, it was shown that the critical point satisfies the following equation:
\begin{eqnarray}\label{eq.BiancPcGenerUncorr}
\frac{\langle k(k-1)\rangle}{\langle k\rangle}
\sum_m\frac{m(m-1)Q(m)}{\langle m\rangle}p_c^{\,m-1}=1.
\end{eqnarray}

\begin{figure}[ht]
  \centering
  \includegraphics[scale=0.45]{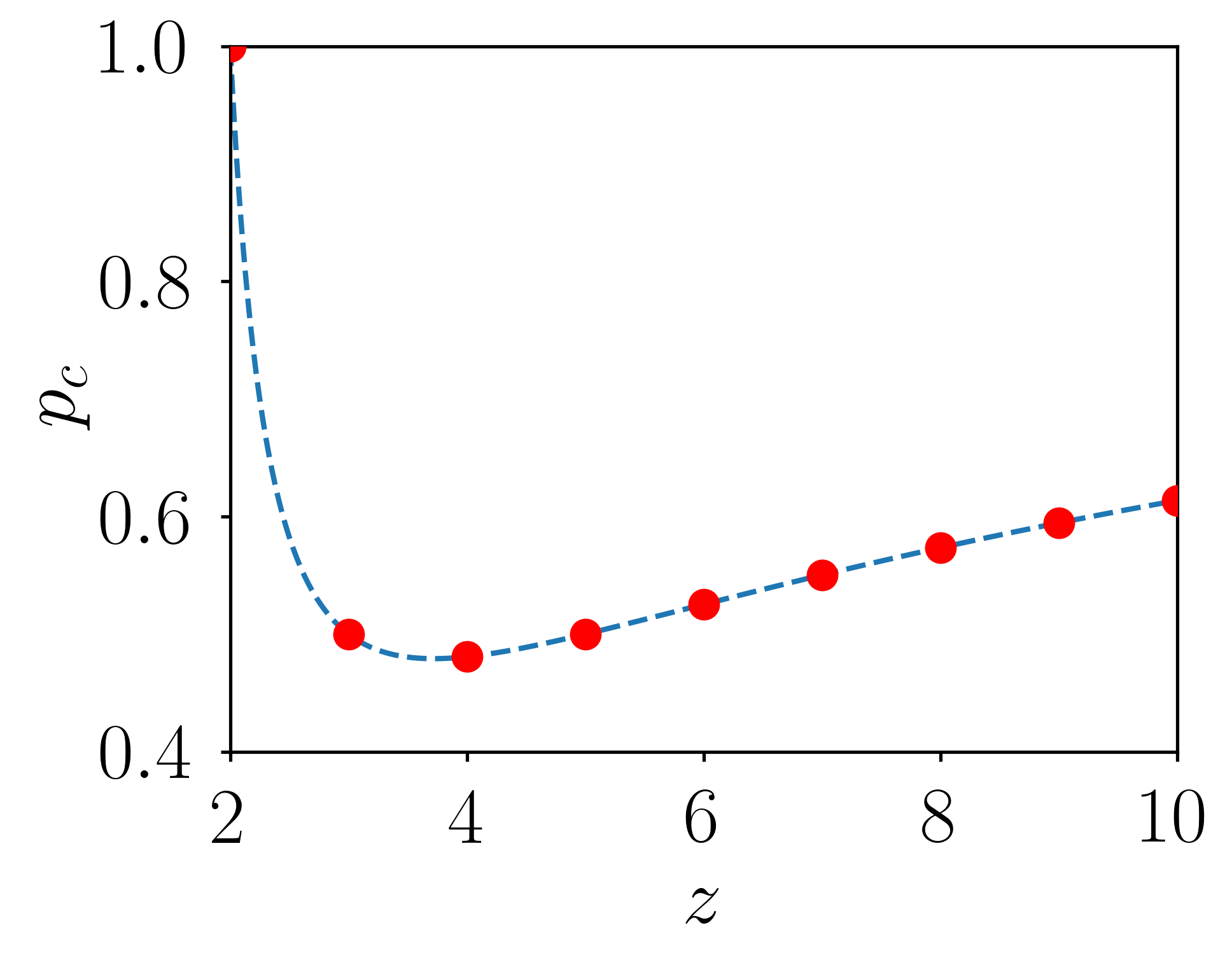}
\caption{Critical percolation threshold $p_c$ as a function of $z$ for random regular hypergraphs in which all nodes have hyperdegree $z$ and all hyperedges have cardinality $z$. Even though $z$ only takes integer values, we show the analytical prediction as a dashed line by extending the expression $p_c = 1/(z-1)^{2/(z-1)}$ to real values of $z$ in order to better visualize the trend. The red dots indicate the actual values of $p_c$ for integer $z$. The threshold exhibits a nonmonotonic dependence on $z$, with a minimum at $z=4$, indicating that the most robust configuration is achieved at this value. For $z>4$, $p_c$ increases (or, in other words, the robustness decreases) as hyperedges become larger and are therefore more easily destroyed by the removal of a single member node. On the other hand, for $z<4$, nodes participate in only a few hyperedges, which also limits the formation of a GC.}
\label{fig.pcRegularHyper}
\end{figure}

To build some intuition before tackling the general correlated case, it is instructive to look at a particularly simple example. Consider a hypergraph where every node has exactly the same hyperdegree $z$ and every hyperedge has exactly the same cardinality $z$, namely, $P(k)=\delta_{k,z}$ and $Q(m)=\delta_{m,z}$. In other words, every node participates in exactly $z$ hyperedges, and each hyperedge contains exactly $z$ nodes. We will refer to this class of networks as \textit{random regular hypergraphs}. For this particular case, the percolation threshold formula [predicted by Eq.~(\ref{eq.BiancPcGenerUncorr})] simplifies to a clean closed-form expression:
\begin{eqnarray}\label{eq.RRHyper}
p_c = \frac{1}{(z-1)^{2/(z-1)}}.
\end{eqnarray}
In Fig.~\ref{fig.pcRegularHyper}, we plot $p_c$ as a function of $z$. Interestingly, the robustness of the hypergraph does not vary monotonically with $z$, but instead displays an optimal value around $z=4$, where $p_c \approx 0.48$. To understand why, consider what happens at the two extremes. For large $z$, each hyperedge contains many nodes, which means that removing just one node is enough to destroy the entire hyperedge. This makes high-$z$ hypergraphs particularly vulnerable. On the other hand, for small values of $z$ (such as $z=2$ or $z=3$), the network is also relatively fragile because nodes participate in only a few hyperedges, which makes the formation of a GC more difficult. As a consequence, the competition between these two effects leads to a maximum robustness at an intermediate value of $z$, which corresponds to $z=4$.

In what follows, we will generalize the equations proposed by Bianconi and Dorogovtsev to account for the effect of correlations between node hyperdegree and hyperedge cardinality, and study how such correlations modify the percolation threshold $p_c$.

\subsection{Theory for correlated hypergraphs}

Let us consider a factor graph (or hypergraph) characterized by a hyperdegree distribution $P(k)$ and a cardinality distribution $Q(m)$, with corresponding excess-hyperdegree and excess-cardinality distributions $P_1(\widetilde{k})$ and $Q_1(\widetilde{m})$, respectively. We define $P_1(\widetilde{k},\widetilde{m})$ as the joint excess hyperdegree-cardinality distribution, which represents the probability that a node with excess hyperdegree $\widetilde{k}$ is connected to a factor node with excess cardinality $\widetilde{m}$.

We assume that $k\in [k_{\min},k_{\max}]$, where $k_{\min}$ and $k_{\max}$ denote the minimum and maximum hyperdegree, respectively; and $m\in[m_{\min},m_{\max}]$, where $m_{\min}$ and $m_{\max}$ denote the minimum and maximum hyperedge cardinality, respectively. The correlation between hyperdegree and cardinality is measured by the usual Pearson coefficient given by:
\begin{equation}\label{eq.corrkCard}
 r=\frac{\sum_{\widetilde{k},\widetilde{m}} \widetilde{k} \widetilde{m} [P_1(\widetilde{k},\widetilde{m}) - P_1(\widetilde{k})Q_1(\widetilde{m})]}{\sigma_{\widetilde{k}} \sigma_{\widetilde{m}}},
\end{equation}
where $\sigma_{\widetilde{k}} = \sqrt{ \sum_{\widetilde{k}} \left(\widetilde{k}\right)^2 P_1(\widetilde{k}) - \left[\sum_{\widetilde{k}} \widetilde{k} P_1(\widetilde{k})\right]^2}$ and $\sigma_{\widetilde{m}} = \sqrt{ \sum_{\widetilde{m}} \widetilde{m}^2 Q_1(\widetilde{m}) - \left[\sum_{\widetilde{m}} \widetilde{m} Q_1(\widetilde{m})\right]^2}$.

The case $r > 0$ (assortative) describes a topology where nodes belonging to many groups (high $k$) tend to participate in groups with large cardinality, while nodes with few connections prefer smaller groups. This could, for example, represent a scenario in which sociable people interact with each other in many and large groups, while more reserved individuals prefer to interact with each other in fewer and smaller groups. Conversely, the case $r<0$, corresponding to disassortative mixing, describes a scenario in which nodes with large hyperdegree tend to belong to small groups, while nodes with small hyperdegree participate in groups with large cardinality.

We now consider the same percolation process as before, in which a fraction $1-p$ of nodes is removed, along with all their hyperedges, but now in the presence of these correlations. Extending the self-consistent equations proposed by Bianconi and Dorogovtsev~\cite{bianconi2024theory} to the case in which node hyperdegree and hyperedge cardinality are correlated, we obtain the following system:
\begin{eqnarray}
W_{\widetilde{m}}&=&\sum_{\widetilde{k}}P_{1}(\widetilde{k}|\widetilde{m})(1-(1-V_{\widetilde{k}})^{\widetilde{k}}),\\
V_{\widetilde{k}}&=&\sum_{\widetilde{m}}P_{1}(\widetilde{m}|\widetilde{k})p^{\widetilde{m}}(1-(1-W_{\widetilde{m}})^{\widetilde{m}}),
\end{eqnarray}
where $\widetilde{k}\in [k_{\min}-1,k_{\max}-1]$, $\widetilde{m}\in[m_{\min}-1,m_{\max}-1]$, $W_{\widetilde{m}}$ is the probability that a randomly chosen node, reached by following a link emanating from a factor node of cardinality $m = \widetilde{m}+1$, belongs to the GC. Similarly, $V_{\widetilde{k}}$ is the probability that a randomly chosen factor node, reached by following a link emanating from a node of hyperdegree $k = \widetilde{k}+1$, is connected to the GC.

The fraction of nodes belonging to the GC is then given by
\begin{eqnarray}\label{eq.PinfHyper}
P_{\infty}&=&p\sum_{k}P(k)\left(1-(1-V_{k-1})^k\right).
\end{eqnarray}
On the other hand, the critical point corresponds to the value of $p$ such that the branching matrix:
\begin{eqnarray}\label{eq.branchCardpc}
B_{\widetilde{k}\widetilde{k'}} = \sum_{\widetilde{m}}p^{\widetilde{m}} \widetilde{k'}P_1(\widetilde{k'}|\widetilde{m})\cdot\widetilde{m}P_1(\widetilde{m}|\widetilde{k}),
\end{eqnarray}
has its largest eigenvalue equal to 1.

In what follows, we will consider a simple but illustrative example in order to understand how these correlations affect the robustness of the system.

\subsection{Trimodal hypergraphs with Degree-Cardinality Correlations}\label{sec.HypTri}
To investigate the effect of degree–cardinality correlations in hypergraphs,  we begin with a simplified model in which both the excess hyperdegree distribution $P_1(\widetilde{k})$  and the excess cardinality distribution $Q_1(\widetilde{m})$ take only three possible values:
\begin{eqnarray}
P_1(\widetilde{k}) &=& P_1(\widetilde{k}_1)\delta_{\widetilde{k},\widetilde{k}_1} + P_1(\widetilde{k}_2)\delta_{\widetilde{k},\widetilde{k}_2} + P_1(\widetilde{k}_3)\delta_{\widetilde{k},\widetilde{k}_3}, \\
Q_1(\widetilde{m}) &=& Q_1(\widetilde{m}_1)\delta_{\widetilde{m},\widetilde{m}_1} + Q_1(\widetilde{m}_2)\delta_{\widetilde{m},\widetilde{m}_2} + Q_1(\widetilde{m}_3)\delta_{\widetilde{m},\widetilde{m}_3}.
\end{eqnarray}
Due to the large number of degrees of freedom in the general case, we restrict our analysis to the symmetric case $\widetilde{k}_i = \widetilde{m}_i$ and $P_1(\widetilde{k}_i) = Q_1(\widetilde{m}_i) = \pi_i$ with $i=1,2,3$. This means that the hyperdegree and cardinality distributions have the same shape.

For this hypergraph, we consider that the degree–cardinality correlations are described by the matrix:
\begin{equation}\label{eq.Matkm}
P_1(\widetilde{k},\widetilde{m})=\begin{pmatrix}
\pi_1-\alpha_1-\alpha_2& \alpha_1 & \alpha_2 \\
\alpha_1 & \pi_2-2\alpha_1 & \alpha_1 \\
\alpha_2 & \alpha_1 & \pi_3-\alpha_1-\alpha_2
\end{pmatrix},
\end{equation}
which satisfies the marginal constraints $\sum_{\widetilde{m}} P_1(\widetilde{k}_i,\widetilde{m})=\pi_i$ and $\sum_{\widetilde{k}} P_1(\widetilde{k},\widetilde{m}_i)=\pi_i$. Just as in the dyadic trimodal case, the two free parameters $\alpha_1$ and $\alpha_2$ allow us to explore a wide range of correlations while keeping both $P(k)$ and $Q(m)$ fixed. In what follows, we will present our results for the specific trimodal hypergraph with $k_1=m_1=2$, $k_2=m_2=4$, $k_3=m_3=7$, and $\pi_1=0.25$, $\pi_2=0.5$, $\pi_3=0.25$. Additional results for other trimodal hypergraphs are presented in Appendix~\ref{appHyper}.

In Fig.~\ref{fig.hyperTriPhase}a, we show the degree-cardinality correlation predicted by Eq.~(\ref{eq.corrkCard}). From this figure, we observe that the behavior of the correlation coefficient is qualitatively similar to what we found for trimodal dyadic networks (see Fig.~\ref{fig.corrPlane}a), which is expected since the joint probability matrix has the same functional form in both cases. Additionally, as in the dyadic case, this parametrization allows us to explore a wide range of correlations ($-0.96 \lesssim r \le 1$). However, despite this similarity in the correlation structure, the effect of those correlations on network robustness turns out to be markedly different from the dyadic case, as we will discuss below. 

In Fig.~\ref{fig.hyperTriPhase}b, we show the critical percolation threshold $p_c$ in the $\alpha_1$-$\alpha_2$ plane. Unlike ordinary dyadic networks, where assortative mixing tends to increase robustness, we find that positively correlated hypergraphs are significantly more fragile than disassortative hypergraphs. For example, in the most disassortative configuration  ($\alpha_1=0$, and $\alpha_2=0.25$) we obtain $p_c \approx 0.166$, which corresponds to the lowest value in the parameter space. On the other hand, for the maximum positive correlation ($\alpha_1=\alpha_2=0$), we get $p_c \approx 0.4807$, which is close to the threshold of a random regular hypergraph with $z=4$ (see Fig.~\ref{fig.pcRegularHyper}).

\begin{figure}[ht]
  \subfigure[]{
    \begin{minipage}{.48\columnwidth}
      \centering
      \includegraphics[scale=0.40]{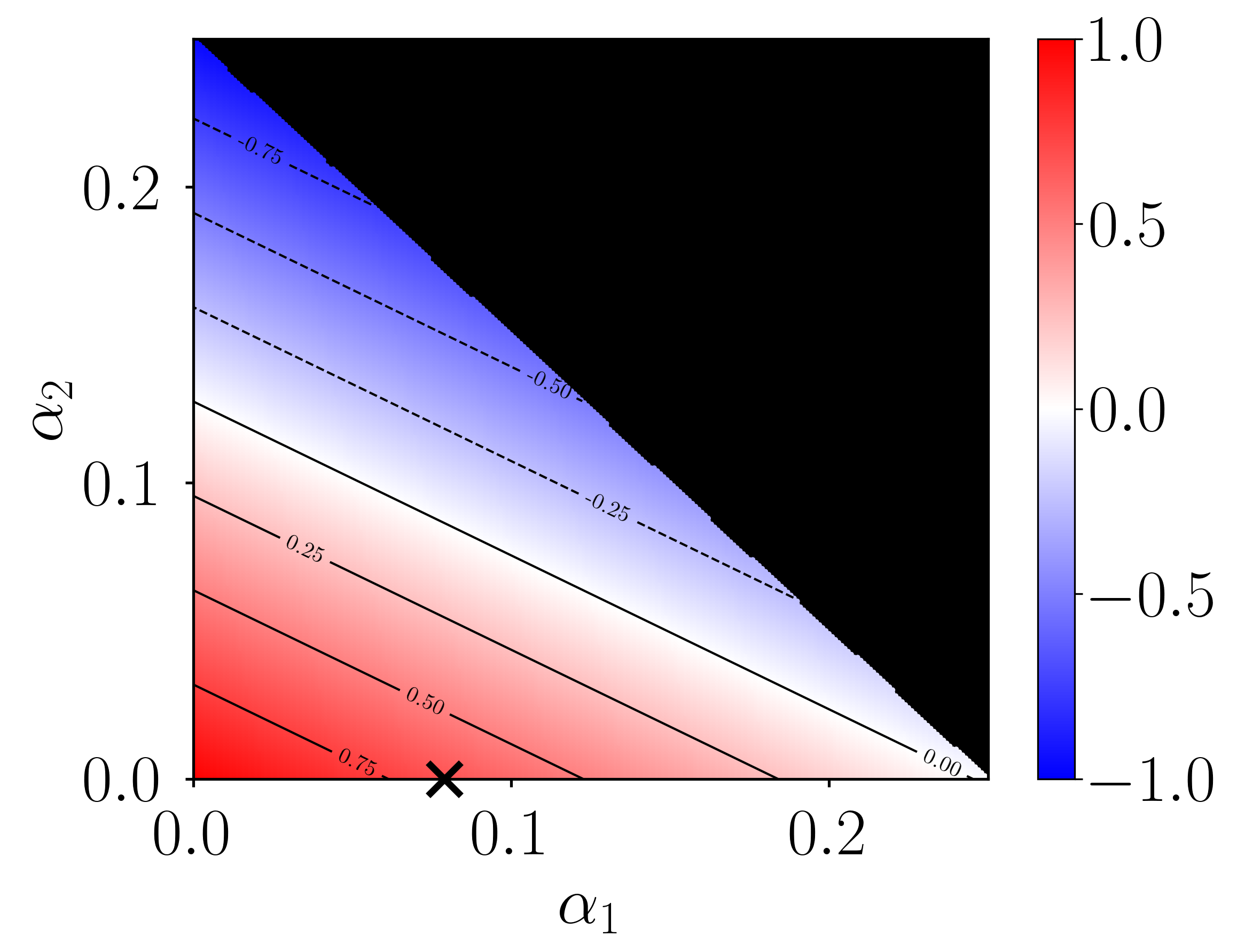}
    \end{minipage}
  }\hfill
  \subfigure[]{
    \begin{minipage}{.48\columnwidth}
      \centering
      \includegraphics[scale=0.40]{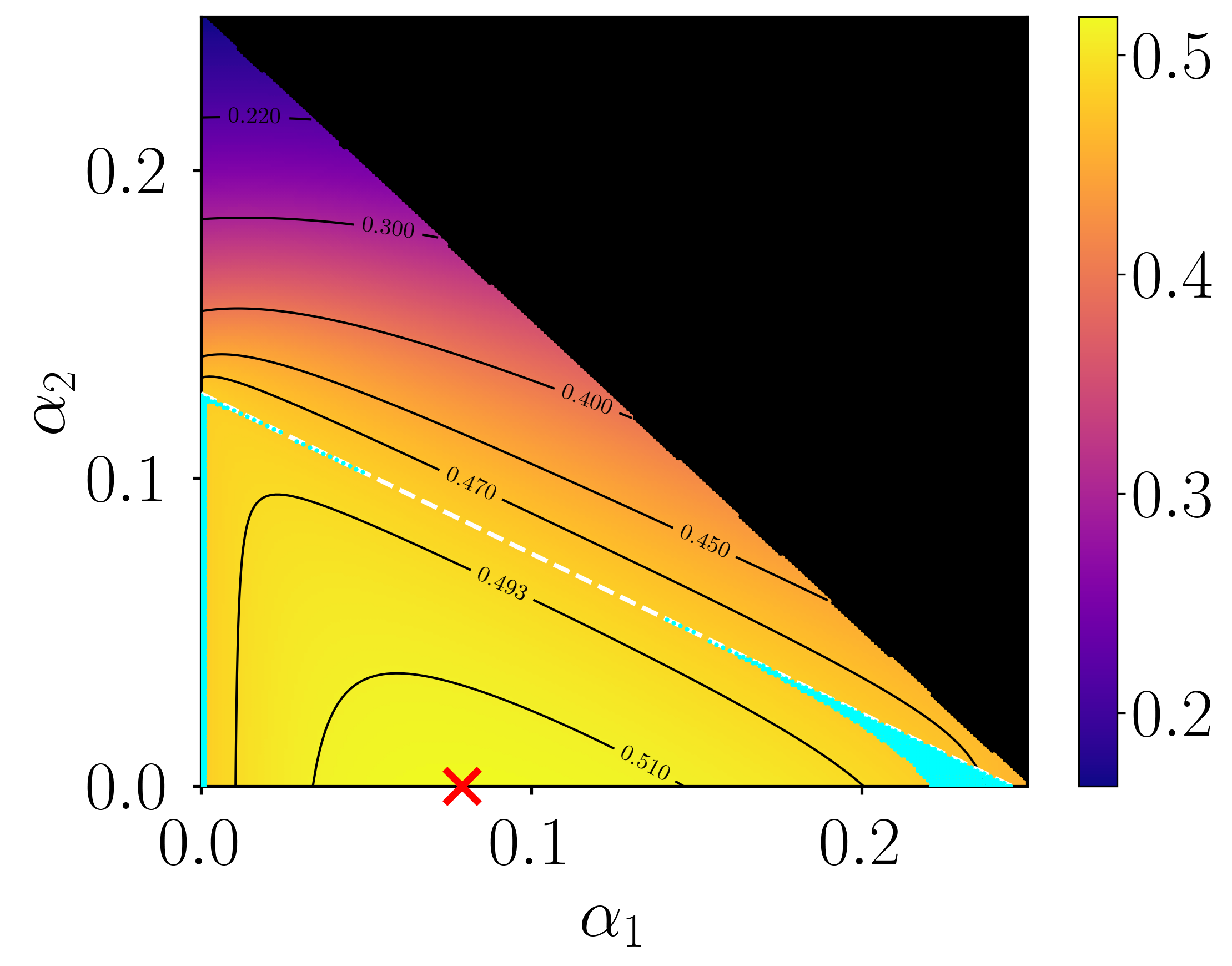}
    \end{minipage}
  }
\caption{Panel (a): Heatmap of the degree–cardinality correlation coefficient $r$ in the $\alpha_1$–$\alpha_2$ plane for the trimodal hypergraph defined in Sec.~\ref{sec.HypTri}. The correlation is computed from Eq.~(\ref{eq.corrkCard}). Straight lines correspond to constant values of $r$.
Panel (b): Heatmap of the critical percolation threshold $p_c$ in the $\alpha_1$–$\alpha_2$ plane obtained from the condition that the largest eigenvalue of the branching matrix [see Eq.~(\ref{eq.branchCardpc})] is equal to 1. The cross in both panels indicates the point of maximum fragility (largest value of $p_c$), which occurs at an intermediate positive correlation. The dashed white line corresponds to the uncorrelated case ($r=0$). The light-blue region indicates the values of $\alpha_1$ and $\alpha_2$ for which assortative hypergraphs exhibit a lower percolation threshold than the fully uncorrelated case, defined by the factorized distribution $P_1(\widetilde{k},\widetilde{m})=P_1(\widetilde{k})Q_1(\widetilde{m})$.}
\label{fig.hyperTriPhase}
\end{figure}

The reason why hypergraphs with $r>0$ are more fragile than those with $r<0$ can be understood from the structural role played by hyperedges.  In assortative hypergraphs, nodes with large hyperdegree tend to belong to hyperedges with large cardinality. Such hyperedges are particularly vulnerable, since removing a single node is sufficient to destroy the entire hyperedge. As a result, high-hyperdegree nodes are likely to lose many of their group memberships, weakening the overall connectivity.

In disassortative hypergraphs, by contrast, nodes with large hyperdegree tend to participate in many hyperedges of small cardinality. This configuration is structurally more robust because smaller hyperedges are less vulnerable to node removal, and as a consequence, high-hyperdegree nodes are unlikely to lose all their hyperedges simultaneously, preserving the GC under random failures.

Although these results clearly show that assortativity tends to make hypergraphs more fragile, this does not imply a monotonic relationship between $r$ and $p_c$. Indeed, as shown in Fig.~\ref{fig.hyperTriPhase}b, the point of maximum fragility corresponds to a positive but intermediate correlation $r \approx 0.68$ ($\alpha_1=0.079$, $\alpha_2=0$), where $p_c \approx 0.52$. Thus, similarly to what was observed in dyadic networks, the most fragile configuration does not coincide with the extreme correlation limit.

\begin{figure}[H]
  \subfigure[]{
    \begin{minipage}{.48\columnwidth}
      \centering
      \includegraphics[scale=0.45]{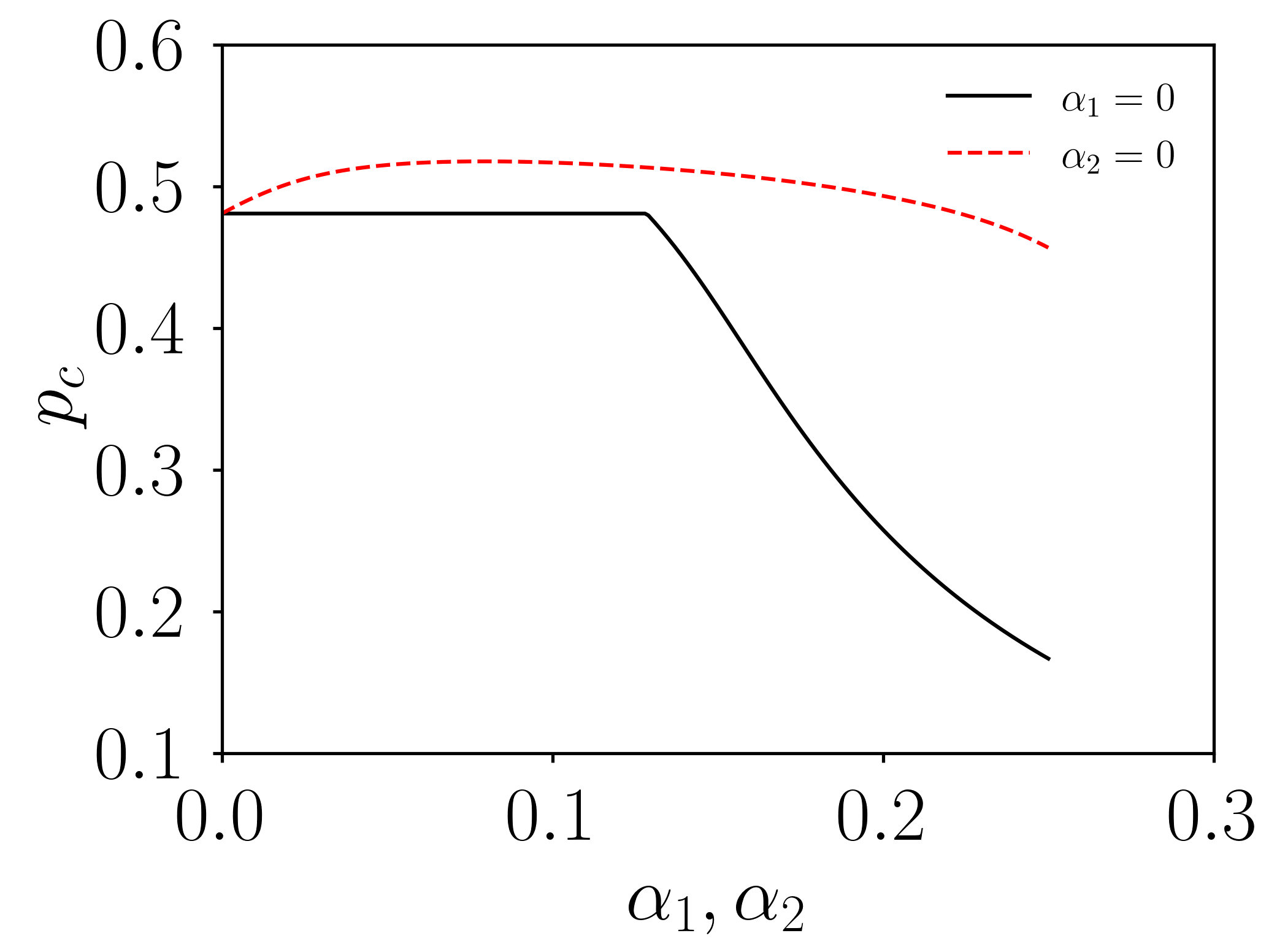}
    \end{minipage}
  }\hfill
  \subfigure[]{
    \begin{minipage}{.48\columnwidth}
      \centering
      \includegraphics[scale=0.49]{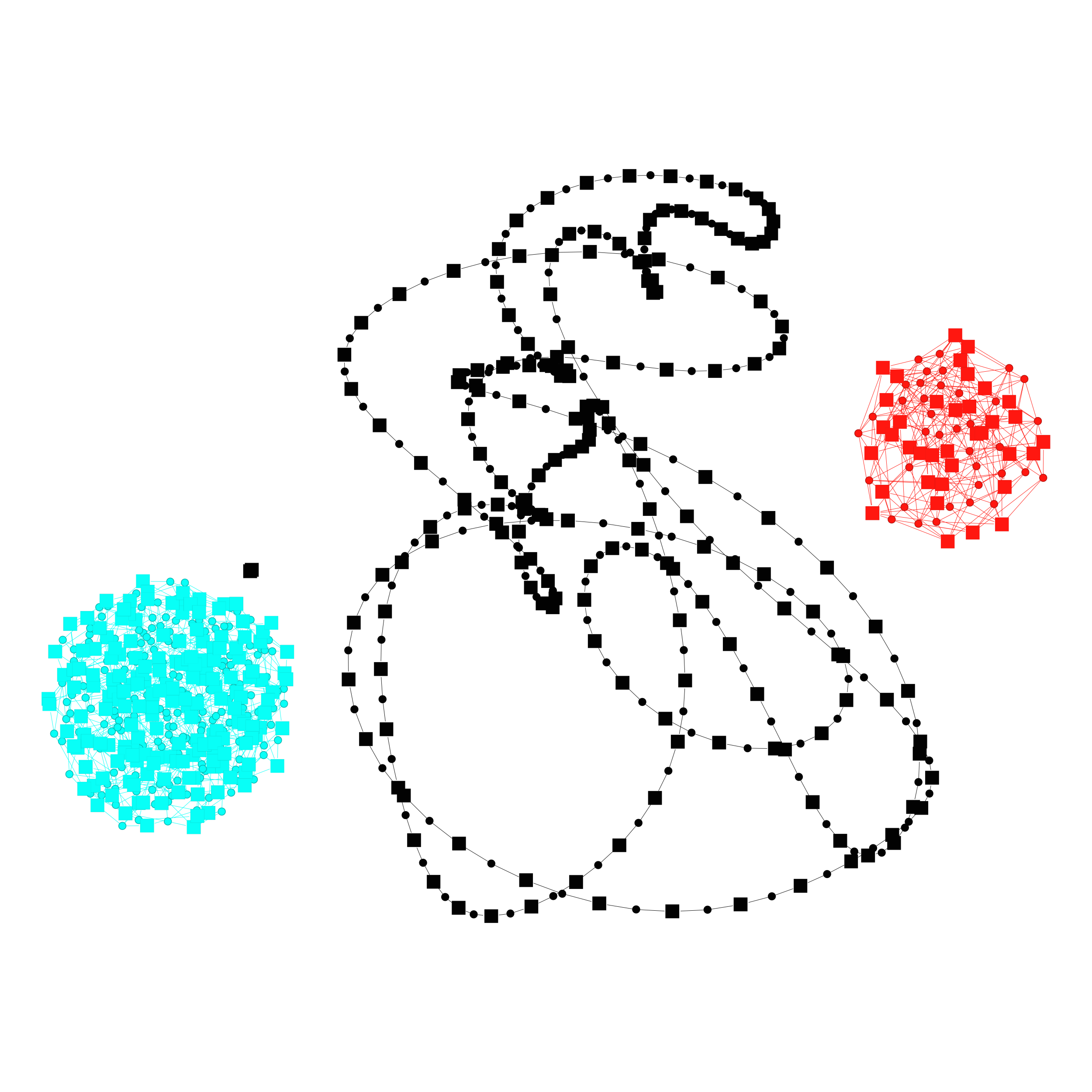}
    \end{minipage}
  }\\
  \subfigure[]{
    \begin{minipage}{.48\columnwidth}
      \centering
      \includegraphics[scale=0.69]{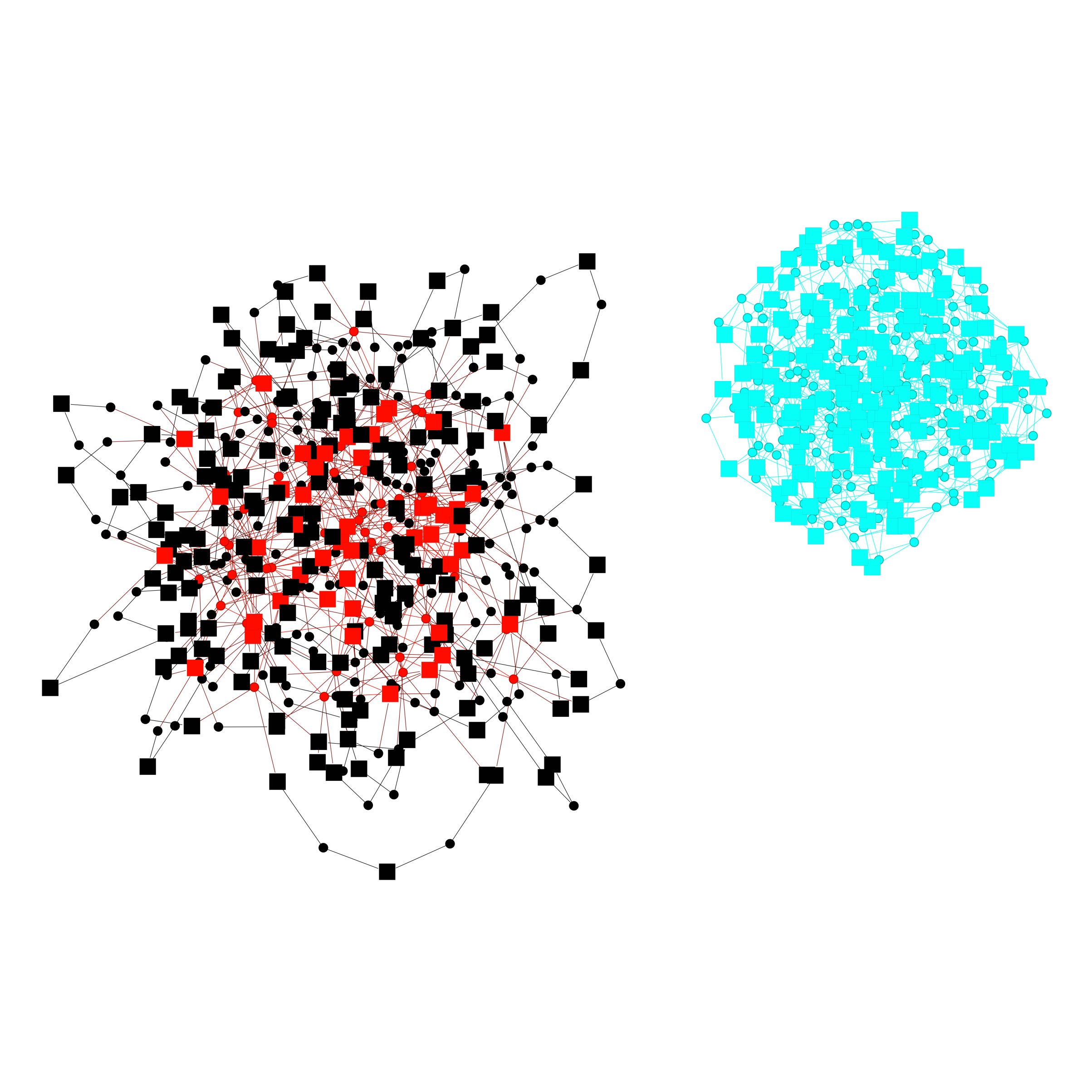}
    \end{minipage}
  }\hfill
  \subfigure[]{
    \begin{minipage}{.48\columnwidth}
      \centering
      \includegraphics[scale=0.69]{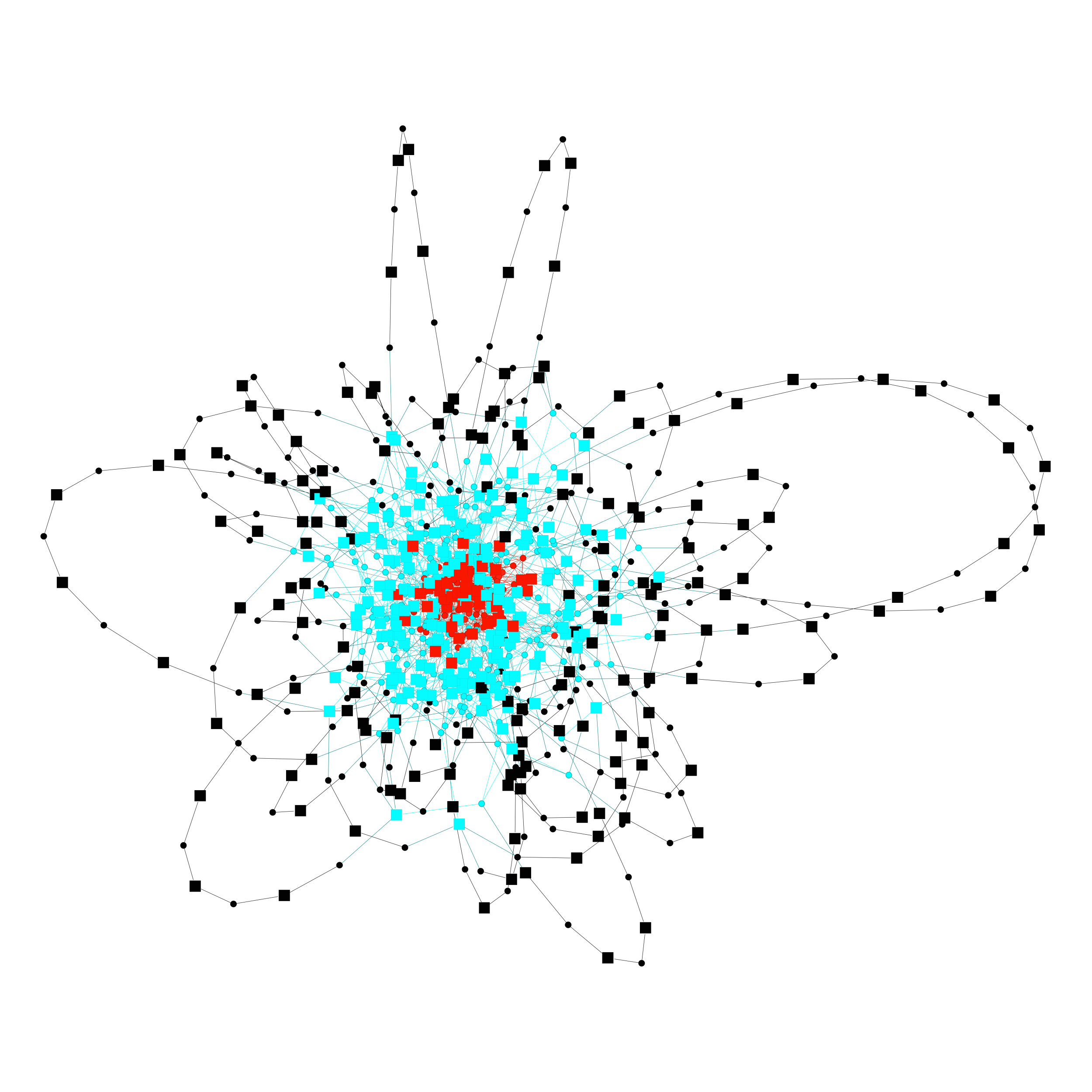}
    \end{minipage}
  }
\caption{Panel (a): Critical threshold $p_c$ along two different paths in the $\alpha_1-\alpha_2$ plane for the trimodal hypergraph defined in Sec.~\ref{sec.HypTri}. The solid line corresponds to the path $\alpha_1=0$ with increasing $\alpha_2$, while the dashed line corresponds to the path $\alpha_2=0$ with increasing $\alpha_1$. Both curves represent theoretical predictions obtained from Eq.~(\ref{eq.branchCardpc}). Panels (b)-(d): Visualizations of the trimodal hypergraph generated using Gephi~\cite{bastian2009gephi} for three different parameter choices: $\alpha_1=\alpha_2=0$ (panel b), $\alpha_1=0$ and $\alpha_2=0.15$ (panel c), and $\alpha_1=0.079$ and $\alpha_2=0$ (panel d). Circles represent nodes and squares represent factor nodes, both colored according to their hyperdegree/cardinality: black ($k=m=2$), light blue ($k=m=4$), and red ($k=m=7$). In panel c, the intermediate-degree community ($k=m=4$) remains completely segregated from the rest of the network, consistent with the fact that $\alpha_1=0$ prevents any mixing between the $k=m=4$ nodes and the other degree classes. In panel d, intermediate-degree nodes begin to mix with both low- and high-degree nodes and factor nodes, disrupting the isolated community structure observed in panel b.}
\label{fig.pcPaths}
\end{figure}

In order to understand the origin of the nonmonotonic behavior, it is useful to compare two particular paths in parameter space $\alpha_1-\alpha_2$: (i) the case $\alpha_1=0$ with increasing $\alpha_2$, and (ii) the case $\alpha_2=0$ with increasing $\alpha_1$. In both cases, the correlation decreases, but the effect on the percolation threshold $p_c$ is qualitatively different, as illustrated in Figs.~\ref{fig.pcPaths}a-d.

In Fig.~\ref{fig.pcPaths}b, we show the maximally assortative case $\alpha_1=\alpha_2=0$, where the hypergraph is fully segregated into three disconnected communities, each characterized by a single value of $z=k=m$. In this configuration, each community behaves as an independent random regular hypergraph.  Among these three communities, the one with $z=4$ has the lowest percolation threshold, and therefore sets the global value of $p_c$. On the other hand, in Fig.~\ref{fig.pcPaths}c, we show the case $\alpha_1=0$ and $\alpha_2=0.15$. Compared with panel (b), the intermediate-degree community with $k=m=4$ remains fully segregated from the rest of the system, so its internal percolation properties are essentially unchanged. More generally, this community remains isolated as $\alpha_2$ increases while $\alpha_1=0$ (see the second row of Eq.~(\ref{eq.Matkm})). At the same time, the low- and high-degree sectors are no longer completely separated as $\alpha_2$ increases: nodes with large hyperdegree ($k=7$) begin to connect preferentially to factor nodes with small cardinality ($m=2$). Since smaller hyperedges are less vulnerable to node removal, these new connections tend to make that part of the structure more robust.

Therefore, by comparing panels (b) and (c), we see that increasing $\alpha_2$ while keeping $\alpha_1=0$ does not make the hypergraph more fragile. The robust $k=m=4$ community is preserved, while the rest of the structure becomes, if anything, more stable. This helps us understand why the critical threshold $p_c$ does not increase along this path, as shown in Fig.~\ref{fig.pcPaths}a.

In contrast, along the second path (i.e., $\alpha_2=0$ and increasing $\alpha_1$) the $k=m=4$ substructure becomes progressively mixed with both the $k=m=2$ and $k=m=7$ components, as can be seen in Fig.~\ref{fig.pcPaths}d, which corresponds to $\alpha_2=0$ and $\alpha_1=0.079$.  As $\alpha_1$ increases, nodes with hyperdegree $k=4$ start connecting to factor nodes of small cardinality $m=2$, which tends to increase robustness. However, these same nodes also begin connecting to factor nodes of large cardinality $m=7$, which are more vulnerable to node removal. These two opposing tendencies may therefore contribute to the nonmonotonic dependence of $p_c$ on $\alpha_1$ along this path, as shown in Fig.~\ref{fig.pcPaths}a.

Finally, another result of the nontrivial relationship between correlations and robustness is illustrated in Fig.~\ref{fig.hyperTriPhase}b. In analogy with the results reported by Wang et al.~\cite{wang2022basic} for dyadic networks, we observe that there exists a broad region of the parameter space in which assortative hypergraphs are more robust than the uncorrelated case, whose critical threshold is given by Eq.~(\ref{eq.BiancPcGenerUncorr}). Combined with the results discussed above for disassortative hypergraphs, this suggests that, in a region around $r=0$, both assortative and disassortative hypergraphs can be more robust than the uncorrelated case, in a way reminiscent of the behavior reported by Hasegawa et al.~\cite{hasegawa2012robustness} for the GC of susceptible nodes. Similar results are shown in Figs.~\ref{fig.AddihyperTriPhase}a-d for other hypergraphs in Appendix~\ref{appHyper}.

\subsection{Percolation in Correlated Hypergraphs with General Degree Distributions}

In the previous section, we saw that degree–cardinality correlations can have a strong effect on the robustness of trimodal hypergraphs. Here, we will extend our analysis to hypergraphs with more general hyperdegree and cardinality distributions. To generate correlated structures while keeping both $P(k)$ and $Q(m)$ fixed, we use the IPF procedure introduced earlier for dyadic networks.

Recall that we denote by $P_1(\widetilde{k})$ and $Q_1(\widetilde{m})$ the excess hyperdegree and excess cardinality distributions, respectively. We first define the cumulative distributions
\begin{equation}
F_k(\widetilde{k}) = \sum_{\widetilde{j} \le \widetilde{k}} P_1(\widetilde{j}),
\end{equation}
\begin{equation}
F_m(\widetilde{m}) = \sum_{\widetilde{j} \le \widetilde{m}} Q_1(\widetilde{j}),
\end{equation}
which correspond to the cumulative distribution functions associated with $P_1(\widetilde{k})$ and $Q_1(\widetilde{m})$, respectively.

Following the same copula-based construction used in the dyadic case, we define an initial seed matrix $P_1^{(0)}(\widetilde{k},\widetilde{m})$ as:
\begin{eqnarray}
P_1^{(0)}(\widetilde{k},\widetilde{m}) &=& C_{\theta}\left(F_k(\widetilde{k}),F_m(\widetilde{m})\right)
- C_{\theta}\left(F_k(\widetilde{k}-1),F_m(\widetilde{m})\right) \nonumber \\
&& - C_{\theta}\left(F_k(\widetilde{k}),F_m(\widetilde{m}-1)\right)
+ C_{\theta}\left(F_k(\widetilde{k}-1),F_m(\widetilde{m}-1)\right).
\end{eqnarray}

It is important to note that, in contrast with the dyadic case, this matrix is not necessarily symmetric because the distributions $P(k)$ and $Q(m)$ can be different. Moreover, the matrix doesn't need to be square, since the number of possible values of $\widetilde{k}$ and $\widetilde{m}$ may also differ.

Starting from this initial matrix, we apply Steps~2 and~3 of the IPF algorithm iteratively until the error between the current marginals and the target distributions $P_1(\widetilde{k})$ and $Q_1(\widetilde{m})$ falls below $10^{-10}$. Once the final joint distribution $P_1(\widetilde{k},
\widetilde{m})$ is obtained, we compute the percolation threshold $p_c$ from the condition that the largest eigenvalue of the branching matrix equals~1 [see Eq.~(\ref{eq.branchCardpc})].

We consider the following five cases:
\begin{itemize}
\item Case I (symmetric): $P(k)$ and $Q(m)$ follow a truncated Poisson distribution $\mathrm{Pois}(4,1,20)$.
\item Case II (symmetric): $P(k)$ and $Q(m)$ follow a truncated scale-free distribution $\mathrm{PL}(1.5,2,20)$.
\item Case III (symmetric): $P(k)$ and $Q(m)$ follow a truncated scale-free distribution $\mathrm{PL}(2.5,2,50)$.
\item Case IV (asymmetric): $P(k)$ follows $\mathrm{Pois}(4,1,20)$ while $Q(m)$ follows $\mathrm{Pois}(7,1,20)$.
\item Case V (asymmetric): $P(k)$ follows $\mathrm{Pois}(6,1,20)$ while $Q(m)$ follows $\mathrm{Pois}(2,1,20)$.
\end{itemize}

\begin{figure}[ht]
  \centering
  \includegraphics[scale=0.60]{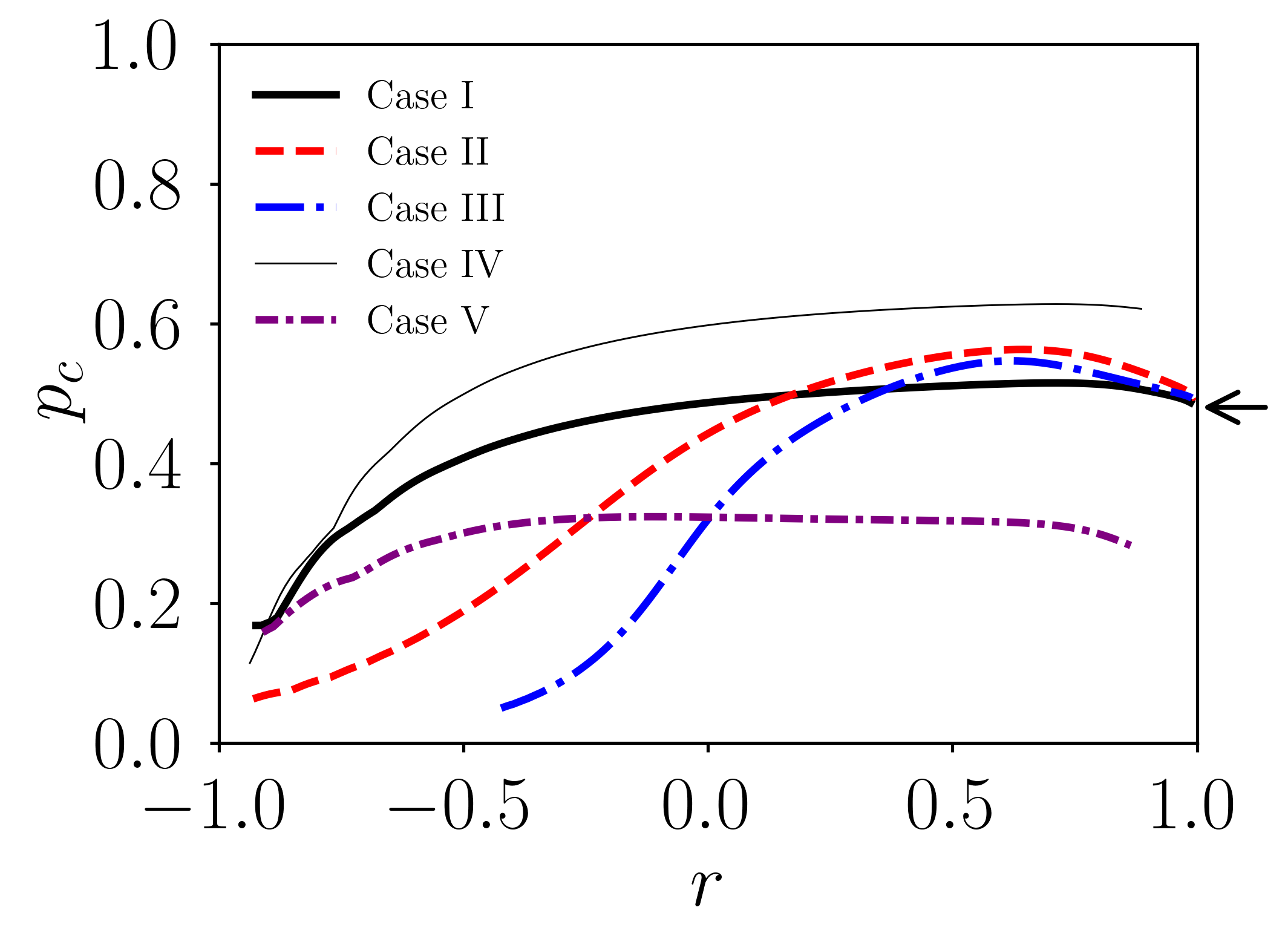}
\caption{Critical percolation threshold $p_c$ as a function of the correlation coefficient $r$ for the five hypergraph configurations described in the text. The arrow indicates the value of $p_c\approx 0.48$, which corresponds to the critical point for a random regular hypergraph with $z=4$ [see Eq.~(\ref{eq.RRHyper})].}
\label{fig.hyperpcvsrCop}
\end{figure}

The results are shown in Fig.~\ref{fig.hyperpcvsrCop}. Across all five cases, we observe that the most negatively correlated configuration corresponds to the most robust hypergraph, i.e., the one with the smallest value of the critical threshold $p_c$. In contrast, assortative hypergraphs exhibit considerably larger values of $p_c$. In particular, for the symmetric cases (Cases I, II, and III), the correlation can approach $r \simeq 1$, and in this limit the critical threshold tends to $p_c \approx 0.48$. This value coincides with the critical threshold of a random regular hypergraph with $z=4$, as predicted by Eq.~(\ref{eq.RRHyper}). Notably, this behavior appears independently of whether $P(k)$ and $Q(m)$ follow a Poisson form or a truncated power-law. This suggests that, in the limit $r \to 1$, the hypergraph decomposes into communities where nodes and factor nodes have matching hyperdegree and cardinality, each behaving as a random regular hypergraph. The overall robustness is then dominated by the $z=4$ community, which is the most 
robust configuration as shown in Fig.~\ref{fig.pcRegularHyper}.

Finally, looking more closely at how $p_c$ varies with the correlation coefficient $r$, we observe that $p_c$ initially increases monotonically with $r$. However, for sufficiently large positive correlations, $p_c$ starts to decrease. This behavior is observed across all cases, although it is less pronounced in Case IV, suggesting that $p_c$ can exhibit a nonmonotonic dependence on $r$ in the assortative regime of hypergraphs only.

\section{Conclusions}\label{sec.Concl}

In this work, we have studied how correlations affect the robustness of complex systems under node percolation, considering both dyadic networks and hypergraphs.

For dyadic trimodal networks, our results show that the relationship between assortativity and robustness is not straightforward. Although assortative networks are generally more robust, we found that disassortative mixing can also enhance robustness in certain regimes. In particular, the percolation threshold exhibits a nonmonotonic dependence on the correlation coefficient, and the most fragile configurations do not necessarily correspond to the most negatively correlated networks. For networks with more general degree distributions, such as truncated Poisson and power-law forms, we observe the same qualitative behavior.

On the other hand, for hypergraphs, we found that the role of correlations is qualitatively different from the dyadic case. Here, assortative correlations tend to increase fragility, while disassortative configurations can significantly improve robustness. Additionally, in the regime of strongly positive correlations, we showed that hypergraphs tend to decompose into communities where nodes and factor nodes have matching hyperdegree and cardinality, each behaving as a random regular hypergraph. In this limit, the overall robustness is dominated by the $z = 4$ community, which corresponds to the most robust configuration among random regular hypergraphs, with a critical threshold $p_c \approx 0.48$. Finally, we found that the critical threshold $p_c$ exhibits a nonmonotonic dependence on the correlation coefficient. However, unlike the dyadic case, this nonmonotonic behavior occurs in the assortative regime, where intermediate positive correlation values can lead to the most fragile configurations.

Overall, our results indicate that correlations play a subtle and nontrivial role in determining the resilience of complex systems. The correlation coefficient by itself is not sufficient to predict robustness under random failures, and more detailed characterizations of the correlation structure are required, such as the full joint distribution $P_1(\widetilde{k},\widetilde{k'})$ and $P_1(\widetilde{k},\widetilde{m})$, or the average nearest neighbor degree $k_{nn}(k)$~\cite{boccaletti2006complex,dorogovtsev2010zero}. We hope that the theoretical tools and results developed here will be useful for future studies on the resilience of systems with higher-order interactions and may also find applications in the design of more robust networks.

\section{Acknowledgments}\label{Sec.Ack}
L. D. V. and C. E. L. wish to thank to UNMdP, Argentina (EXA 1193/24) and CONICET, Argentina, for financial support.

\section{Data availability}
The data that support the findings of this article are openly available~\cite{gith01}.

\appendix

\section{Construction of correlated networks}\label{app.Construct}

One of the algorithms commonly used to construct correlated networks with a prescribed joint degree distribution $P_1(\widetilde{k},\widetilde{k'})$ was proposed by Newman in Ref.~\cite{newman2003mixing}. This method is based on the Metropolis-Hastings algorithm. It starts from an uncorrelated network generated using the configuration model~\cite{molloy1995critical}. Then, pairs of links are selected and rewired with a probability that depends on the target distribution $P_1(\widetilde{k},\widetilde{k'})$. This procedure is iterated until the empirical joint degree distribution converges sufficiently close to the target distribution.

Although this algorithm can, in principle, generate networks with a wide range of correlations, our numerical simulations indicate that its convergence becomes prohibitively slow for strongly disassortative networks.  For this reason, we employ a variant of the configuration model, similar to the one proposed in~\cite{raschke2014copula}, to directly construct correlated random networks.

The procedure is as follows:

\begin{enumerate}
\item Assign to each node a degree $k$ (i.e., number of stubs) drawn from the distribution $P(k)$, with $k_{\min} \leq k \leq k_{\max}$.

\item Randomly generate a pair $(\widetilde{k}, \widetilde{k'})$ according to the joint excess-degree distribution $P_1(\widetilde{k},\widetilde{k'})$, where $\widetilde{k}=k-1$ and $\widetilde{k'}=k'-1$ represent outgoing degrees.

\item Select one stub from a node of degree $k=\widetilde{k}+1$ and another stub from a node of degree $k'=\widetilde{k'}+1$, and connect them to form a link, avoiding self-loops and multiple connections.

\item Repeat steps 2 and 3 until no free stubs remain. If, after a large number of attempts, the remaining stubs cannot be successfully connected, we proceed as follows. If the fraction of unmatched stubs is smaller than $0.01\%$ of the total number of stubs, they are discarded. Otherwise, the construction process is aborted and restarted from step 1.
\end{enumerate}
This algorithm can be straightforwardly extended to construct correlated hypergraphs (or equivalently, bipartite networks) with a prescribed joint distribution $P_1(\widetilde{k},\widetilde{m})$. The implementations of these algorithms are available in our public GitHub repository~\cite{gith01}.

While the algorithm described above can in principle generate networks for any joint distribution $P_1(\widetilde{k},\widetilde{k'})$, it is important to note that the range of achievable Pearson coefficients $r$ is not arbitrary. In particular, it has been shown that the degree distribution itself constrains the possible values of $r$, especially in heterogeneous scale-free networks~\cite{yang2017lower}.

\section{Additional results for dyadic networks}\label{appDyaNet}

\begin{figure}[H]
 \subfigure[]{
    \begin{minipage}{.48\columnwidth}
      \centering
      \includegraphics[scale=0.40]{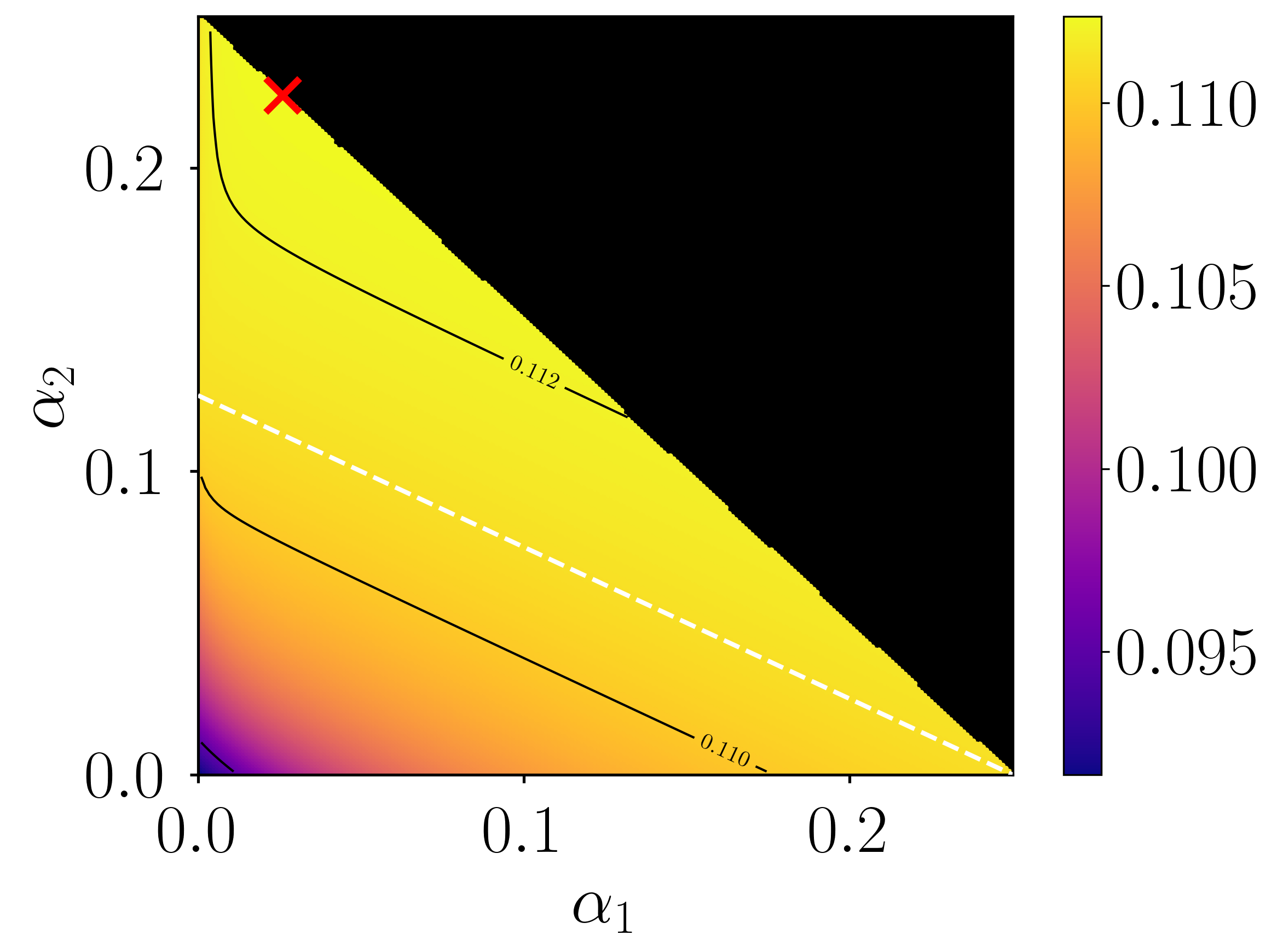}
    \end{minipage}
  }\hfill
  \subfigure[]{
    \begin{minipage}{.48\columnwidth}
      \centering
      \includegraphics[scale=0.40]{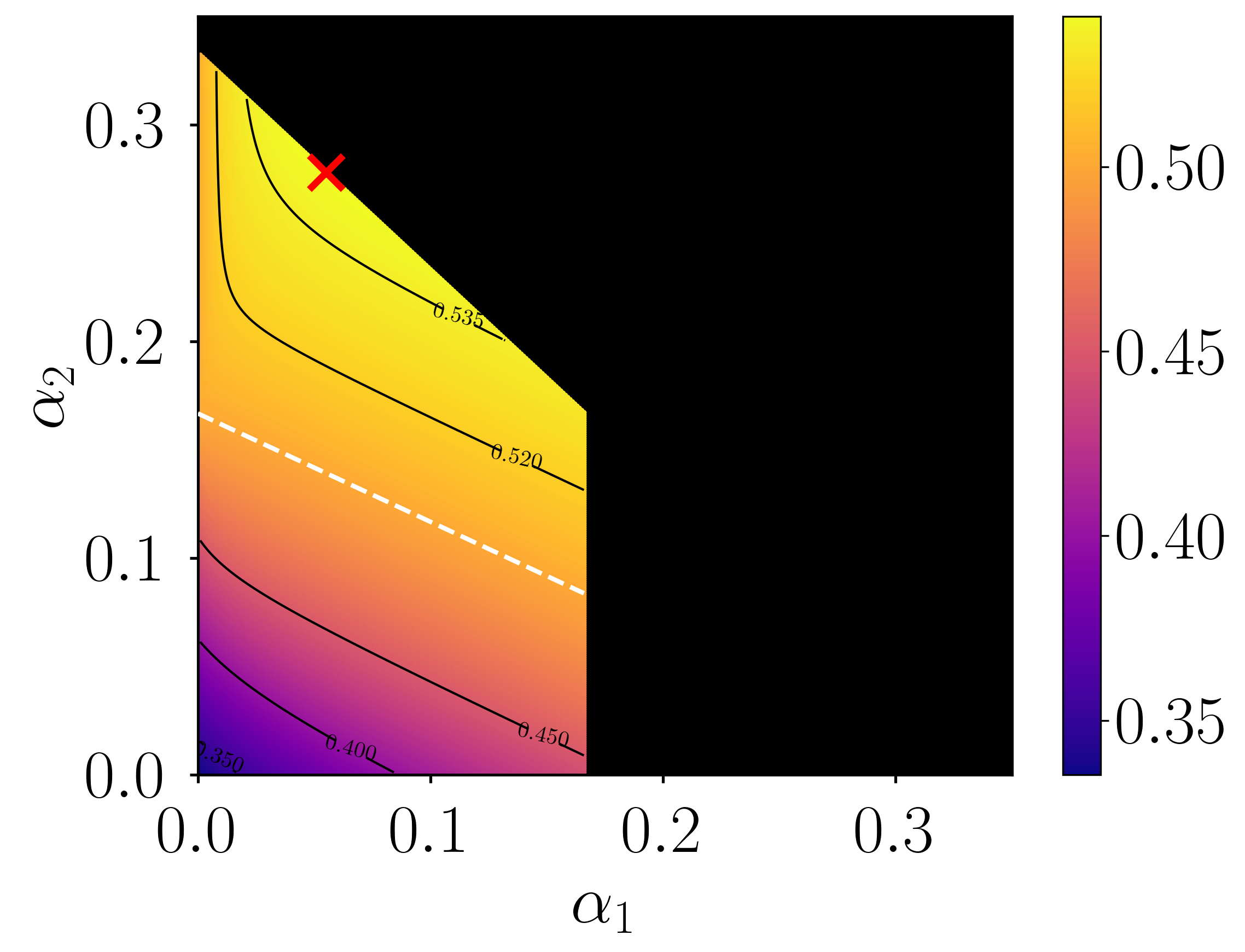}
    \end{minipage}
  }\\
  \subfigure[]{
    \begin{minipage}{.48\columnwidth}
      \centering
      \includegraphics[scale=0.40]{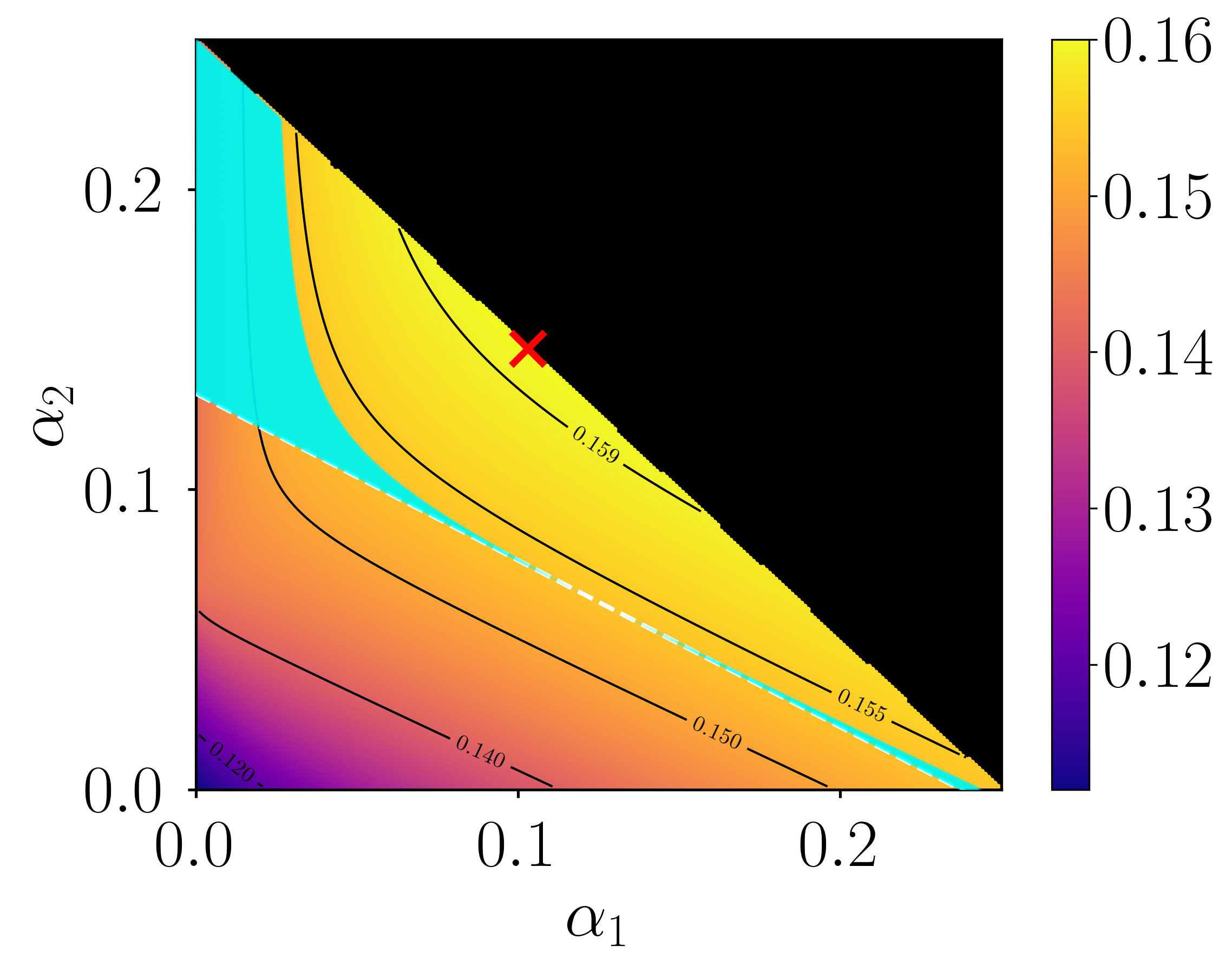}
    \end{minipage}
  }\hfill
  \subfigure[]{
    \begin{minipage}{.48\columnwidth}
      \centering
      \includegraphics[scale=0.40]{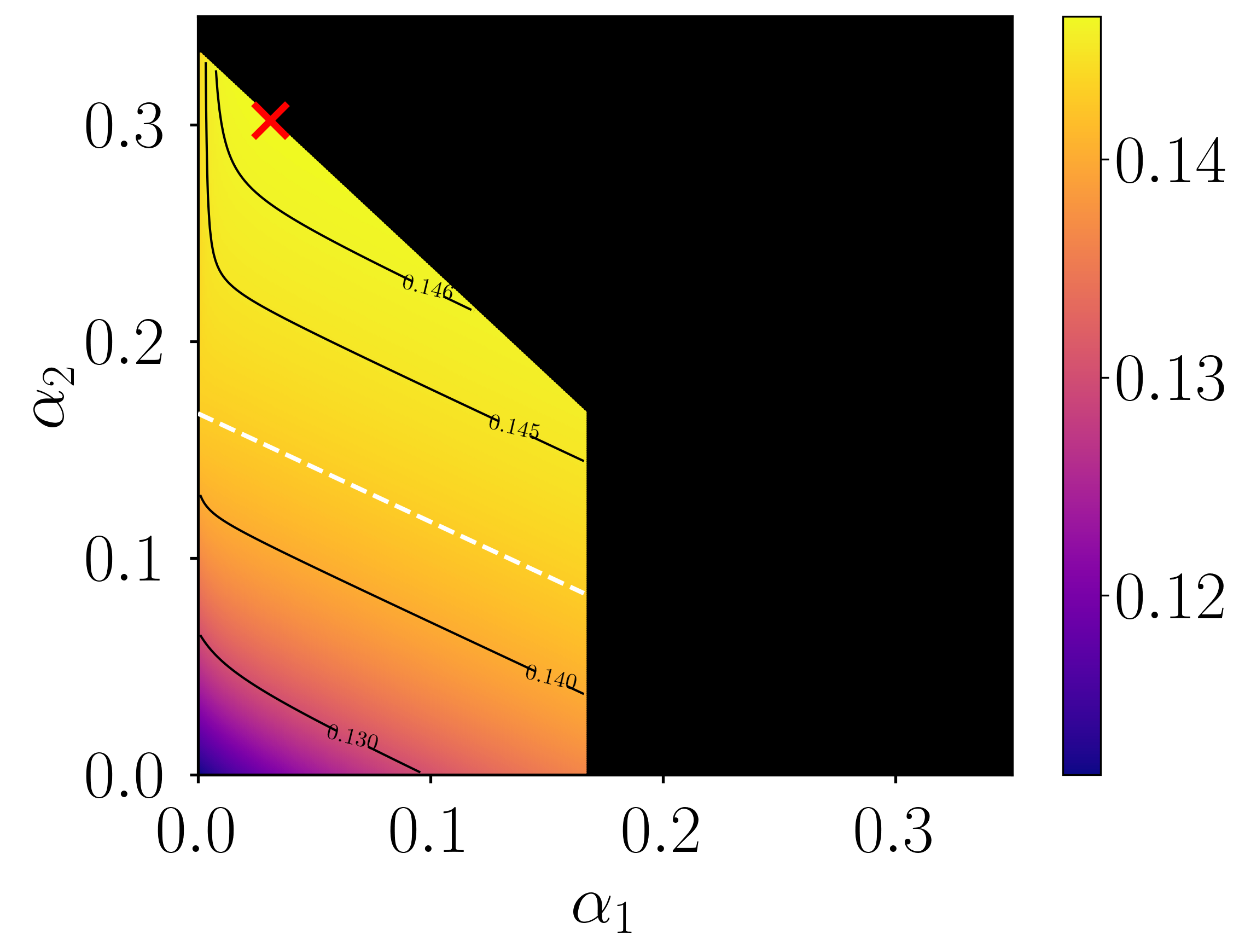}
    \end{minipage}
  }
\caption{Heatmap of the critical percolation threshold $p_c$ in the $\alpha_1$–$\alpha_2$ plane. Panels (a)-(d) correspond to Cases I-IV described in the text, respectively. In all panels, each cross indicates the point of maximum fragility (largest value of $p_c$). The dashed white line corresponds to the uncorrelated case ($r=0$). The light-blue region indicates the values of $\alpha_1$ and $\alpha_2$ for which disassortative networks are more robust than the fully uncorrelated case, defined by the factorized distribution $P_1(\widetilde{k},\widetilde{k'})=P_1(\widetilde{k})P_1(\widetilde{k'})$.}
\label{fig.AddiTriPhase}
\end{figure}

In Fig.~\ref{fig.AddiTriPhase}, we present the critical percolation threshold $p_c$  in the $\alpha_1$-$\alpha_2$ plane for four additional trimodal networks with different degree parameters. Specifically, we consider: 
\begin{itemize}
\item Case I: $k_1=8$, $k_2=10$, $k_3=12$ with $\pi_1=0.25$, $\pi_2=0.50$, $\pi_3=0.25$; 
\item Case II: $k_1=2$, $k_2=3$, $k_3=4$ with $\pi_1=\pi_2=\pi_3=1/3$; 
\item Case III: $k_1=4$, $k_2=8$, $k_3=10$ with $\pi_1=0.25$, $\pi_2=0.50$, $\pi_3=0.25$;
\item Case IV: $k_1=6$, $k_2=8$, $k_3=10$  with $\pi_1=\pi_2=\pi_3=1/3$. 
\end{itemize}
In all cases, we observe that the point of maximum fragility (the largest value of $p_c$) does not coincide with the most disassortative configuration.

\section{Additional results for hypergraphs}~\label{appHyper}

\begin{figure}[htbp]
  \subfigure[]{
    \begin{minipage}{.48\columnwidth}
      \centering
      \includegraphics[scale=0.40]{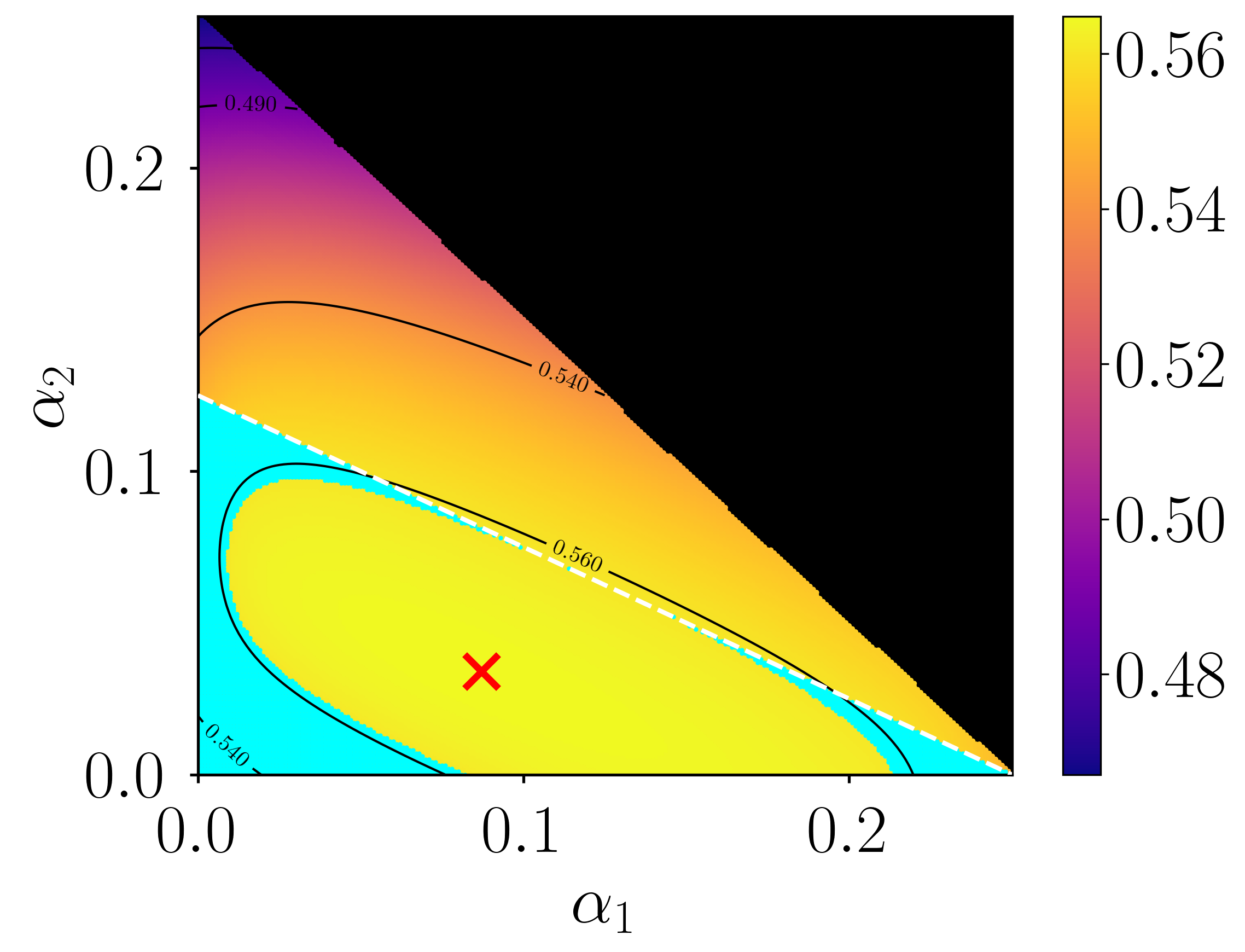}
    \end{minipage}
  }\hfill
  \subfigure[]{
    \begin{minipage}{.48\columnwidth}
      \centering
      \includegraphics[scale=0.40]{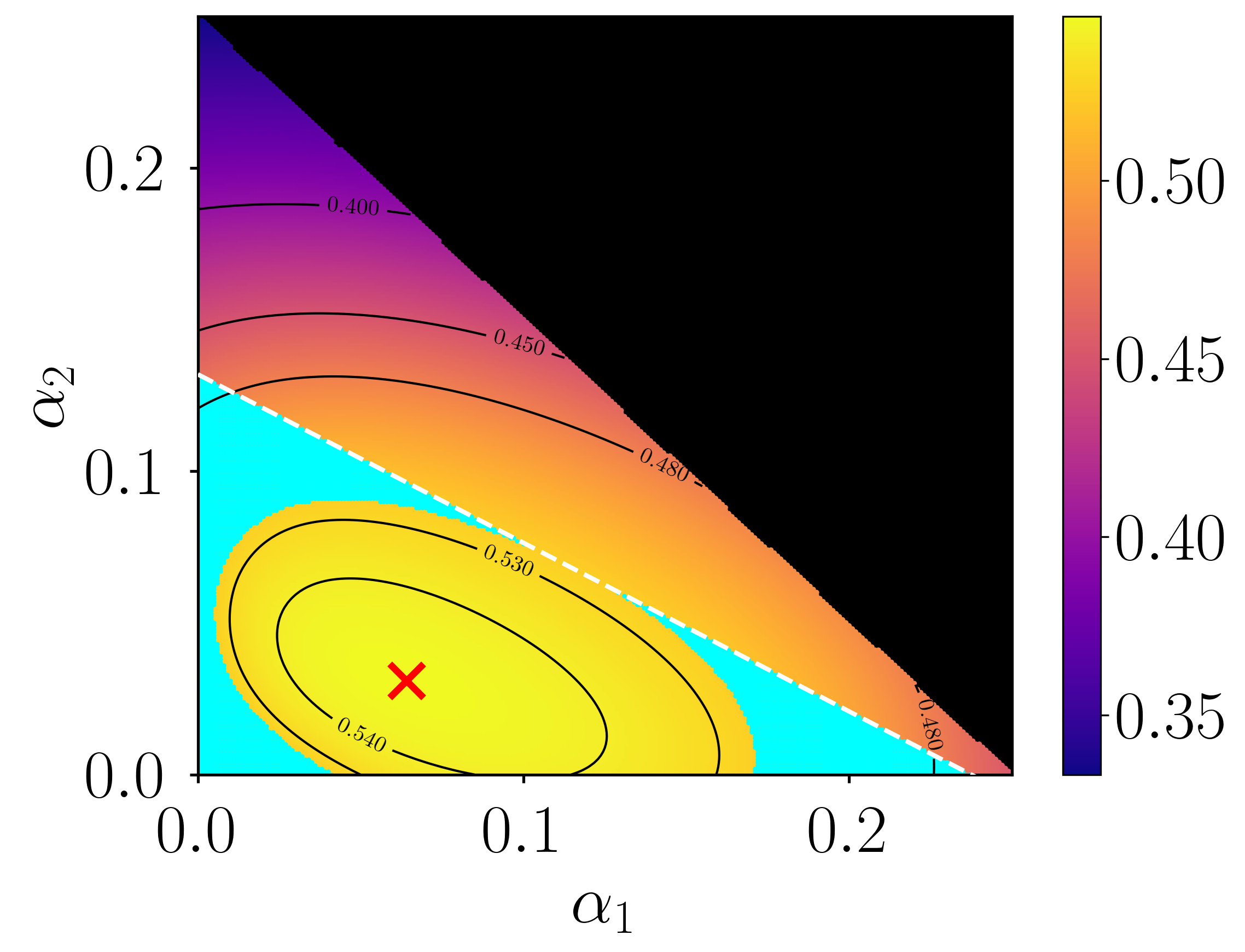}
    \end{minipage}
  }\\
  \subfigure[]{
    \begin{minipage}{.48\columnwidth}
      \centering
      \includegraphics[scale=0.40]{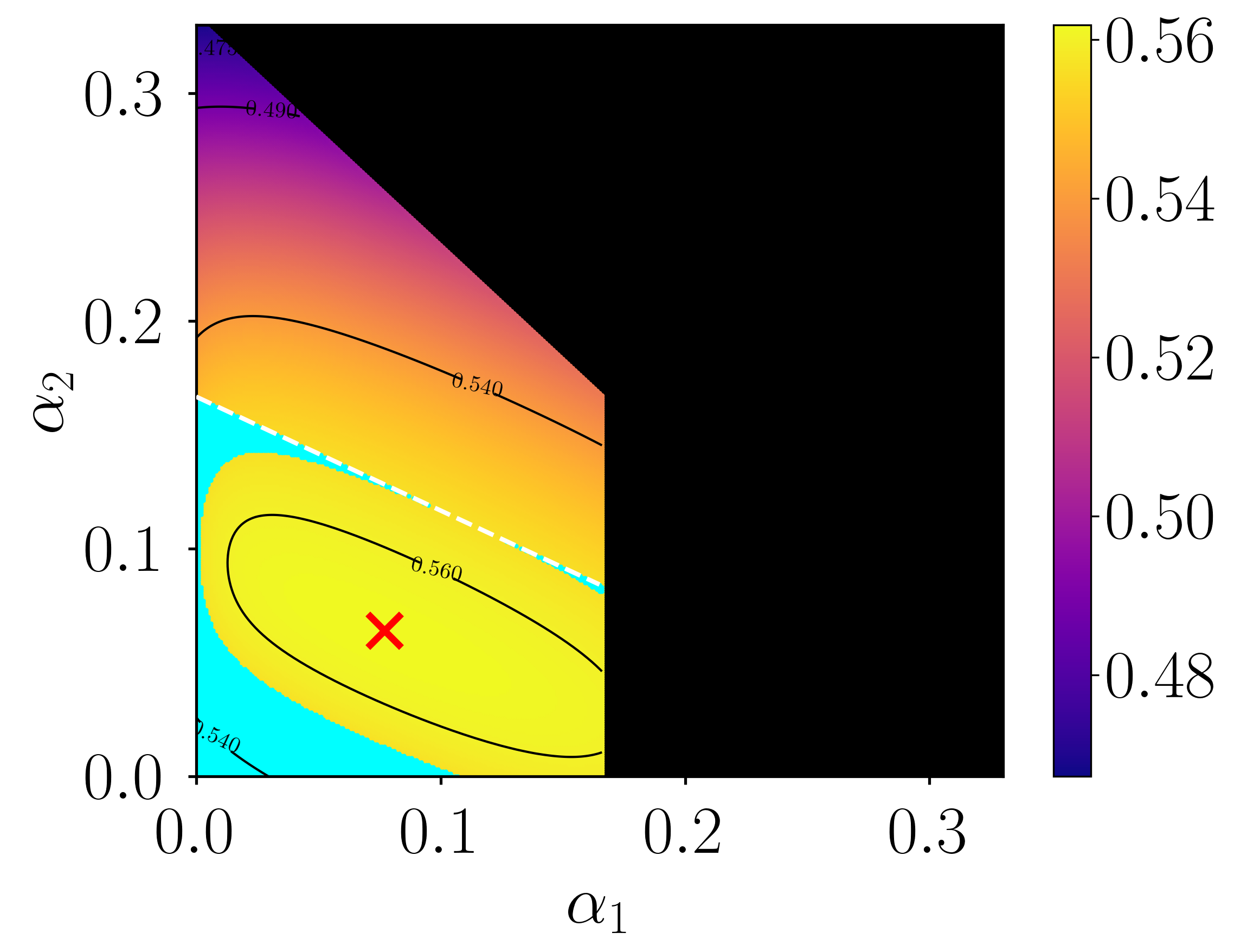}
    \end{minipage}
  }\hfill
  \subfigure[]{
    \begin{minipage}{.48\columnwidth}
      \centering
      \includegraphics[scale=0.40]{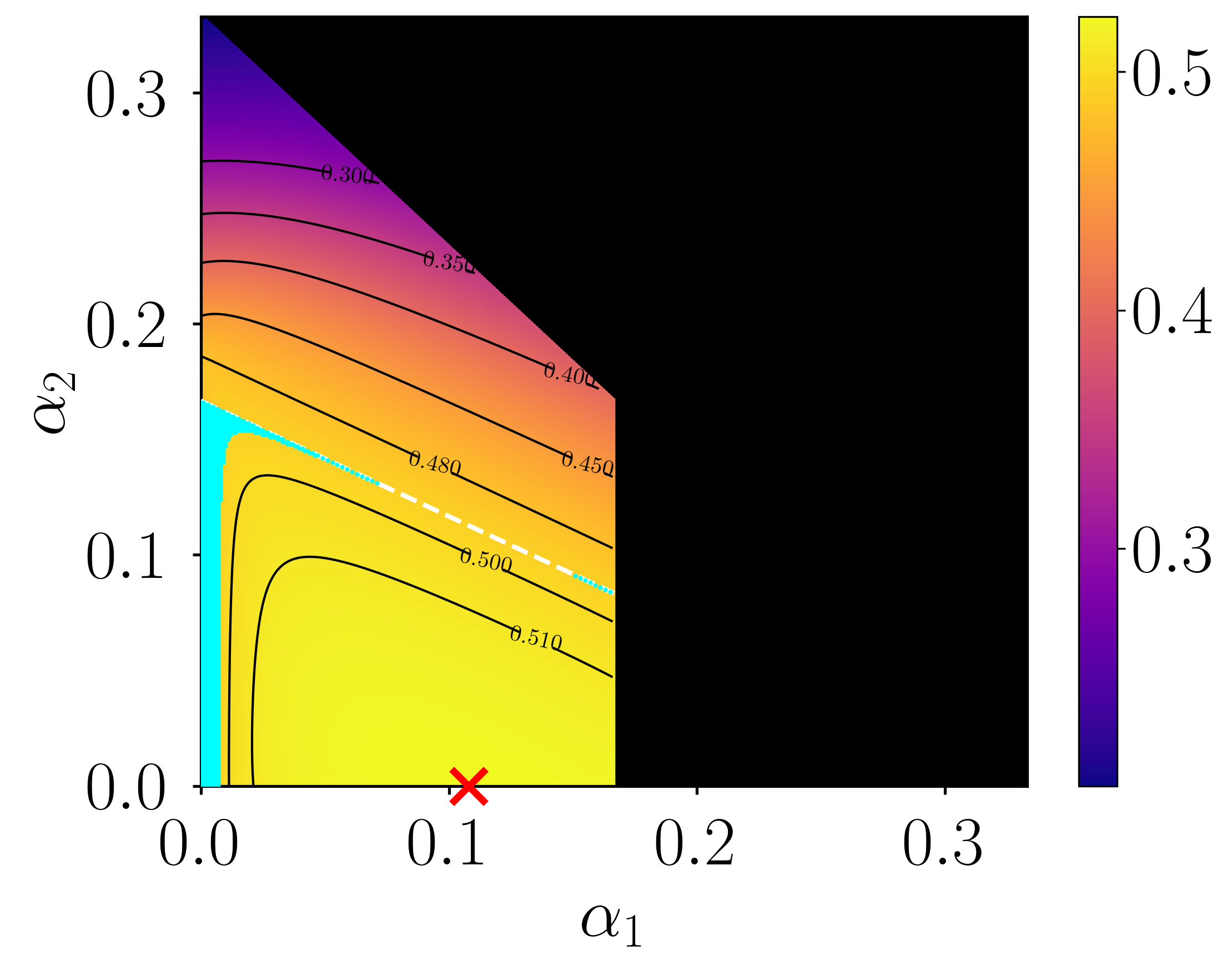}
    \end{minipage}
  }
\caption{Heatmap of the critical percolation threshold $p_c$ in the $\alpha_1$–$\alpha_2$ plane. Panels (a)-(d) correspond to Cases I-IV described in the text, respectively.  Each cross indicates the point of maximum fragility (largest value of $p_c$), which occurs at an intermediate positive correlation. The dashed white line corresponds to the uncorrelated case ($r=0$). The light-blue region indicates the values of $\alpha_1$ and $\alpha_2$ for which assortative hypergraphs exhibit a lower percolation threshold than the fully uncorrelated case, defined by the factorized distribution $P_1(\widetilde{k},\widetilde{m})=P_1(\widetilde{k})Q_1(\widetilde{m})$.}
\label{fig.AddihyperTriPhase}
\end{figure}

In Fig.~\ref{fig.AddihyperTriPhase}, we present the critical percolation threshold $p_c$ in the $\alpha_1$-$\alpha_2$ plane for four additional trimodal hypergraphs, all satisfying the symmetric condition $P(k)=Q(m)$. The specific parameters for each case are:
\begin{itemize}
    \item Case I: $k_1=m_1=6$, $k_2=m_2=8$, $k_3=m_3=10$, 
    $\pi_1=0.25$, $\pi_2=0.50$, $\pi_3=0.25$.
    \item Case II: $k_1=m_1=4$, $k_2=m_2=8$, $k_3=m_3=10$, 
    $\pi_1=0.25$, $\pi_2=0.50$, $\pi_3=0.25$.
    \item Case III: $k_1=m_1=6$, $k_2=m_2=8$, $k_3=m_3=10$, 
    $\pi_1=\pi_2=\pi_3=1/3$.
    \item Case IV: $k_1=m_1=2$, $k_2=m_2=4$, $k_3=m_3=6$, 
    $\pi_1=\pi_2=\pi_3=1/3$.
\end{itemize}
In all cases, we observe that the most fragile configuration corresponds to an assortative hypergraph, and that the point of maximum fragility does not coincide with the most assortative configuration. This is consistent with the nonmonotonic behavior reported in Sec.~\ref{sec.HypTri}.

\begin{figure}[htbp]
  \subfigure[]{
    \begin{minipage}{.48\columnwidth}
      \centering
      \includegraphics[scale=0.45]{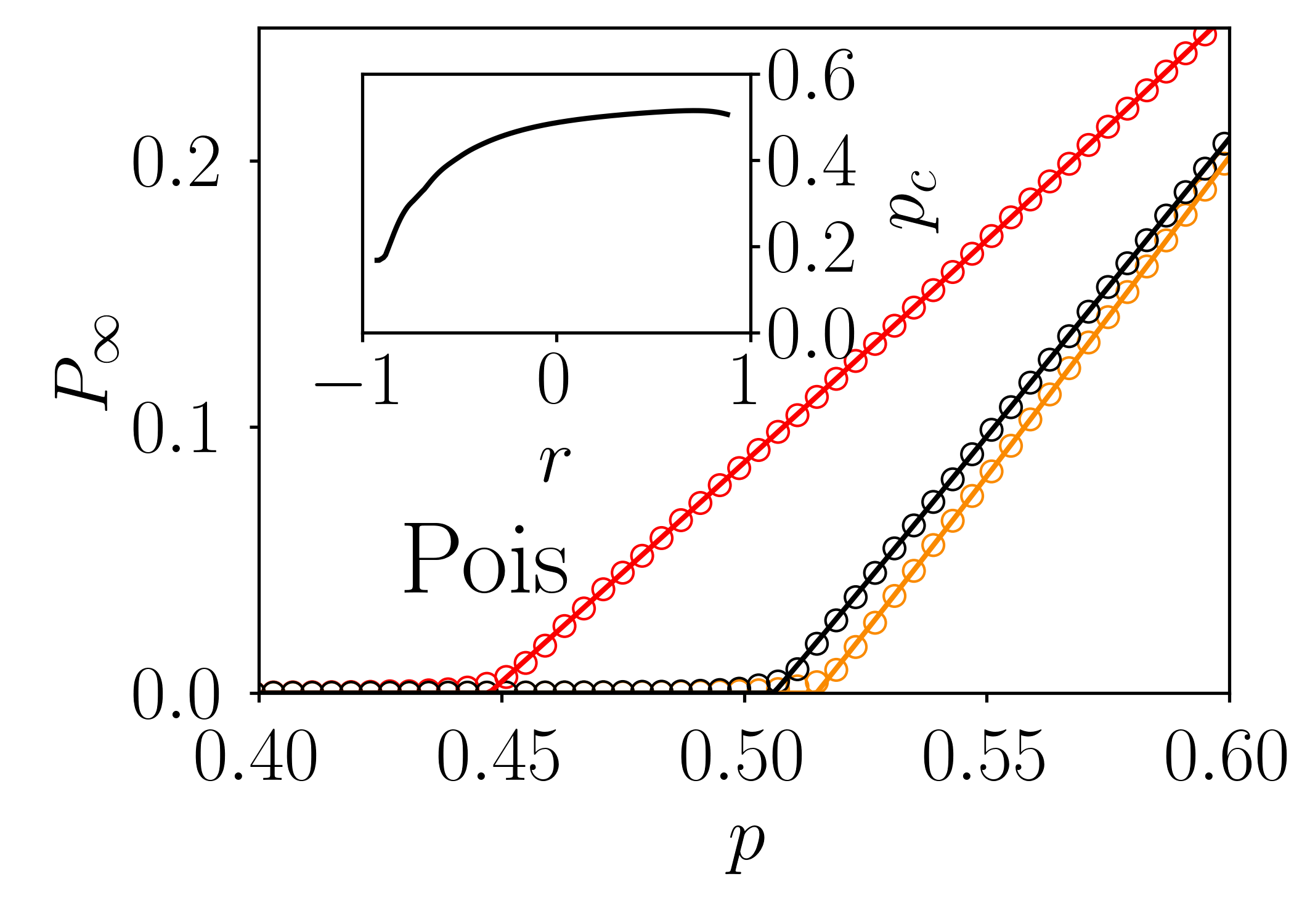}
    \end{minipage}
  }\hfill
  \subfigure[]{
    \begin{minipage}{.48\columnwidth}
      \centering
      \includegraphics[scale=0.45]{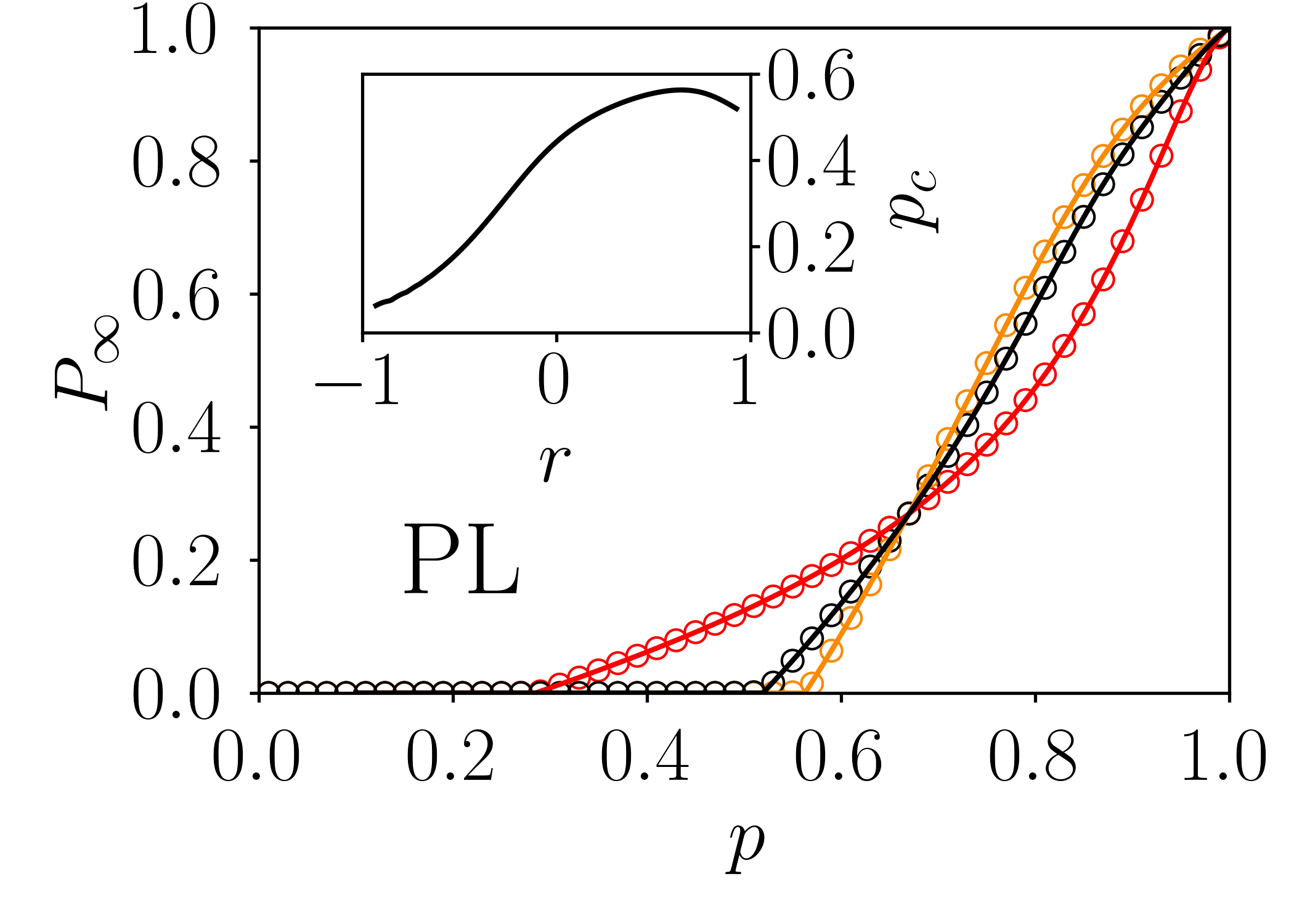}
    \end{minipage}
  }
\caption{Panel (a): In the main panel, we plot $P_{\infty}$ as a function of $p$ for hypergraphs in which  $P(k)$ and $Q(m)$ follow a truncated Poisson distribution $\mathrm{Pois}(4,1,20)$. Solid lines correspond to the theoretical solution of Eq.~(\ref{eq.PinfHyper}), while symbols represent stochastic simulations for networks of size $N=10^6$. Colors indicate different values of the hyperdegree-cardinality correlation coefficient $r$: red $r=-0.34$ (obtained with $\theta=-0.4$ in the Clayton copula), orange $r=0.72$ ($\theta=3.04$), and black $r=0.89$ ($\theta=9$). The inset shows $p_c$ versus $r$ obtained from Eq.~(\ref{eq.branchCardpc}).
Panel (b): Same as in (a) but for hypergraphs in which  $P(k)$ and $Q(m)$ follow a power-law distribution $\mathrm{PL}(1.5,2,20)$. Colors indicate different values of the hyperdegree-cardinality correlation coefficient $r$: red $r=-0.31$ ($\theta=-0.40$), orange $r=0.64$ ($\theta=2.16$), and black $r=0.93$ ($\theta=9$).}
\label{fig.HyperERSF}
\end{figure}

In Fig.~\ref{fig.HyperERSF}, we present additional results for hypergraphs with more general degree distributions. Panel (a) shows $P_{\infty}$ as a function of $p$ for hypergraphs in which both $P(k)$ and $Q(m)$ follow a truncated Poisson distribution $Pois(4,1,20)$, for several values of the correlation coefficient $r$ generated using the Clayton copula (see Sec.~\ref{sec.HyperDegCar}). The inset shows $p_c$ as a function of $r$. Both the main panel and the inset reveal a nonmonotonic dependence of $p_c$ on $r$. Panel (b) shows the same analysis for hypergraphs whose degree and cardinality distributions follow a truncated power-law $PL(1.5,2,20)$, and the same nonmonotonic behavior is observed.

\bibliography{bib}
\end{document}